\documentclass{aa}  
\usepackage{graphicx}
\usepackage{aalongtable}
\usepackage{longtable,lscape}
\usepackage{txfonts}
\begin{document}
   \title{Abundance variations in the globular cluster M71 (NGC 6838)}


   \author{A. Alves-Brito
          \inst{1}
          \and
          R. P. Schiavon\inst{2}
	  \and
	  B. Castilho \inst{3}
	  \and
	  B. Barbuy \inst{1}
          }

   \offprints{A. Alves-Brito}

   \institute{Universidade de S\~ao Paulo, IAG, Rua do Mat\~ao 1226,
Cidade Universit\'aria, S\~ao Paulo 05508-900, Brazil\\
              \email{abrito@astro.iag.usp.br, barbuy@astro.iag.usp.br}
         \and
             Department of Astronomy, University of Virginia, P.O. Box 3818,
	     Charlottesville, VA 22903-0818.\\
             \email{rps7v@mail.astro.virginia.edu}
         \and
              Laborat\'orio Nacional de Astrof\'{\i}sica/MCT, CP 21, Itajub\'a,
	    MG, 37500-000, Brazil\\
             \email{bruno@lna.br}             
             }

   \date{2007}

 
  \abstract
{Abundance variations in moderately metal-rich globular clusters can give
clues about the formation  and chemical enrichment of globular clusters.}   
   {CN, CH, Na, Mg and Al indices in spectra of 89 stars
of the template metal-rich globular cluster M71 are measured and
implications on internal mixing are discussed.}  
   {Stars from the turn-off up to the  Red Giant Branch (0.87
   $<$ log g $<$ 4.65) observed with the GMOS multi-object spectrograph at the
   Gemini-North telescope are analyzed. Radial velocities, colours, 
effective temperatures, gravities
   and spectral indices are determined for the sample.}
   {Previous findings related to the CN bimodality and CN-CH 
anticorrelation in stars of M71 are confirmed. 
We also find a CN-Na correlation,
and Al-Na, as well as an Mg$_2$-Al anticorrelation.}
   {A combination of convective mixing and a primordial pollution by AGB
   or massive stars in the early stages of globular cluster formation is
   required to explain the observations.}


   \maketitle
%

\section{Introduction}

Globular Clusters (GCs)  provide important information on the early
chemical and dynamical evolution of the Milky Way. 

Star-to-star abundance variations of light elements - Li, C, N, O,
Na, Mg, and Al - are extensively reported in the literature (Gratton
et al. 2004 and references therein). Li variations among turnoff
(TO) stars and a Li-Na anticorrelation have been reported; among giant
stars,  C and N abundances, as well as the pairs O:Na and Al:Mg
are also found to be anticorrelated. Such anomalies
have not been obtained for heavier elements.   These abundance
variations have been reported in the literature over the  past
two decades, but their origin is still widely debated.

Some of the abundance variations seen in globular-cluster
giants can be explained by evolutionary mixing with migration of
processed material through the CNO cycle to the surface of giant
stars (Iben 1964; Charbonnel 1994), whereas a primordial-enrichment
scenario, which requires  early contamination of intracluster
material, is claimed by some authors (e.g. Smith 1987; Kraft 1994).

With a moderately high metallicity ($\langle$[Fe/H]$\rangle$=$-$0.73,
Harris 1996), and  an old age of 10-12 Gyr (Grundahl et al.
2002; Meissner \& Weiss 2006), M71 is often regarded as a prototype
of northern metal-rich globular clusters and considered as a suitable
globular cluster to study abundance variations. It is located near
the Galactic plane (b = $-$4.6$^{o}$) and has a reddening of E(B-V)
= 0.27 $\pm$ 0.05 and an apparent distance modulus of (m-M)$_{\rm
V}$ = 13.60 $\pm$ 0.10 (Geffert \& Maintz 2000). Dinescu et al.
(1999) have obtained space velocities for a set of Galactic globular
clusters. For M71 they determined velocity components (U, V, W) =
(77 $\pm$ 14, $-$58 $\pm$ 10, $-$2 $\pm$ 14) km/s and a low
eccentricity orbit,  which characterizes M71, kinematically,
as a thick-disk cluster.

Chemical abundances were discussed in several previous studies of
this cluster, including DDO photometry of giant stars (Hesser
et al. 1977; Briley et al. 2001) and low- and high-resolution
analysis of stars in different evolutionary stages from the
Main-Sequence (MS) TO to the Red Giant Branch (RGB) tip (Cohen 1980;
Smith \& Norris 1982; Leep et al. 1987; Smith \& Penny 1989; Penny
et al.  1992; Sneden et al. 1994; Cohen 1999;  Ramirez \& Cohen
2002; Lee et al. 2004; Lee 2005;  Boesgaard et al. 2005; Yong et
al. 2006).  Many of these studies show a CN bimodality, with CN-CH
anticorrelation, Na-O anticorrelation and  variations in Al abundance.

In this paper we present the main results  of an analysis of
high S/N, medium resolution, Gemini/GMOS spectra of a large number
of M~71 stars, from the main-sequence turnoff to the tip of the
giant branch.  Our goals are twofold: to improve the statistics on
abundance variations in M71 stars; and to study the behaviour of
 14 spectral indices for the sample stars in order to shed
light on the main astrophysical processes leading to the observed
star-to-star abundance variations.   We estimate the C and N
abundances of one CN-strong and one CN-weak giant from spectrum
synthesis, based on state-of-the-art model photospheres and an
up-to-date line list.  A comparison of our results with 
 those based on recent high-resolution abundance
analysis is also presented.


This paper is structured as follows. The observations and data
reductions are described in Sect. 2. The analysis  of radial
velocities, photometry, atmospheric parameters and line indices is
presented in Sect. 3. In Sect. 4 the results are shown, and in Sec.
5 they are discussed and  contrasted with previous studies.
Our conclusions are summarized in Sect. 6.


\section{Observations and data reduction}


Imaging and spectroscopy of M71 stars were obtained with the Gemini
Multi-Object Spectrograph (GMOS; see, for example, Hook et al. 2004
for more details) in the MOS mode on the 8m Frederick C. Gillett
Telescope (Gemini-North).

On July 18, 2002, the pre-imaging required to build four GMOS masks
was obtained. It superposes a GMOS-North field over 5.5 arcmin x
5.5 arcmin. The image of the M71 field was obtained using the
r$_{-}$G0303 filter, with an effective wavelength of 6300 {\rm \AA} and
wavelength coverage of 5620-6980 $\rm \AA$. Exposures of 4 x 180s were
taken, with the CCD detector operating at 4 electrons/DN and a readout
noise of 6.6 electrons. The mean airmass during the pre-imaging
observation was 1.157.  Thus, using a finding-chart available in
Cudworth (1985),  145 M71 stars were identified in the GMOS
image and selected for spectroscopic observations.

Spectroscopic data were collected on August 5, 2002, using
the B600+$_{-}$G5303 (600 lines/mm) grating and adopting
arcsec-wide slits.  The CCDs were binned in a 2 x 2 mode (along
both the spatial and dispersion axes).  This configuration achieved
a spectral resolution R $\sim$ 2,000  at 5100 $\rm \AA$ with
a dispersion of 0.0917 nm/pix, covering from 3500 to 7000 $\rm \AA$.

The 145 target stars were selected in order to give appropriate
sampling of the colour magnitude diagram (CMD).  The maximum time
for each individual exposure was constrained by the bright red
giants of the cluster. Short exposures were taken in order
to avoid saturation and a number of these exposures for the faintest
stars of the turn-off were co-added. The total  integration
time was defined by our desire to achieve S/N $\sim$ 150 at $\lambda$
$\sim$ 4000 $\rm \AA$.  Finally, spectra of the spectrophotometric
standard star EG 131 were obtained using a 1-arcsec longslit, with
the same instrumental set up as adopted for the science observations.

Science and standard star data were both reduced using the GEMINI
GMOS Data Reduction Package  within the IRAF package. Bias
frames, flat-fields and CuAr images were taken as part of the GEMINI
base calibrations.  At this point, the reduction process comprised
bias-subtraction and flat-fielding, using GCAL flats which
were previously co-added and normalized.  Cosmic rays were then
cleaned.  The wavelength calibration was carried out with solutions
obtained from the CuAr arc exposures, which provided typical residuals
of 0.2 $\rm \AA$.  Spectra were then sky-subtracted and extracted
into a series of 1D spectra.  All spectra were  extinction-corrected
using the mean extinction coefficients obtained for Mauna Kea. 
As a result of the location of
the {\it slits} with respect to the  mask-bisector, the
wavelength coverage varies from star to star.  In addition, two
gaps corresponding to 0.5mm between CCDs are also present in the
final spectra.


\section{Analysis}

\subsection{Heliocentric radial velocities}

Heliocentric radial velocities for the sample stars were obtained
using both  {\it rvidlines} and {\it fxcor} IRAF procedures.  The
former measures radial velocities from spectra by determining the
wavelength shift in spectral lines relative to specified rest
wavelengths and the measured velocities are corrected to a heliocentric
frame of reference. We derived a mean heliocentric radial velocity
of $-12$ $\pm$ 46 km/s (N = 145 stars).  The task {\it fxcor}, on
the other hand, performs a Fourier cross-correlation  between
lists of input object and template spectra.  A set of 77 spectral
templates  with stellar parameters given by: [Fe/H] = $-$1.00,
[$\rm \alpha/Fe$]= 0.00, 4250 $\leq$ T$_{\rm eff}$ $\leq$ 5750 K
and 0.00 $\leq$ logg $\leq$ 5.0 dex were taken from the  library
of synthetic stellar spectra by Coelho et al. (2005).  Measurements
of 142 stars led to a mean heliocentric radial velocity of $11$
$\pm$ 49 km/s.

The mean difference between the heliocentric radial velocities as
measured in two ways (v$_{\rm fxcor}^{\rm h}$ - v$_{\rm
rvidlines}^{\rm h}$) is 23 km/s. For comparison, Harris
(1996) gives a mean heliocentric radial velocity of $-$22.8 $\pm$
0.2 km/s for M71.   As the sample spectra have a resolution R
$\sim$ 2000, we expect a theoretical accuracy of around (1/10)(c/2000)
= 15 km/s, which is in good agreement with the scatter obtained
above.  Individually, the higher values in the measurements of the
heliocentric radial velocities could be explained in part based on
the low-resolution of the spectra. We found a nearly Gaussian
distribution of the radial velocitites with a dispersion of $\sim$
45 km/s, and we believe that this finding reflects the uncertainty
in the determination of the radial velocity itself at this spectral
resolution.  The accuracy in the heliocentric radial velocities
measurements is also wavelength-calibration dependent; in our case the GMOS
calibration lamps were rather poor. Given the
high velocity-dispersion obtained, the radial velocities were used
to shift the sample spectra to rest wavelengths rather than a star
membership selection criterion.

\subsection{Photometry}

As reported in the GEMINI/GMOS web
site\footnote{http://www.gemini.edu/sciops/instruments/gmos/gmosIndex.html},
up to September 12, 2006 there was a large ($\sim$5") offset in the
optical positions (RA and DEC) in GMOS-North images.  A FORTRAN
code was applied in order to take these offsets into account and
find out the correct identification of the sample stars from previous
photometric catalogs of M71.  Using the {\it SkyCat} tool
we plotted all 145 observed stars on the Gemini-North preimaging
and created our own finding chart of M71 following the identification
as given in each GMOS mask.  Optical photometry for M71 from the
literature was adopted by selecting CCD photometry  able to give
us crucial information like completeness and membership probability.
Two papers fulfilled these requirements: (i) Cudworth (1985) presents
proper motions and visual photometry for 350 M~71 stars down
to a limit of V = 16.  Cudworth (2006, priv.  comm.)
gives proper motions and photometry for fainter stars (V $>$ 16),
amounting to 1522 stars.  (ii) Geffert \& Maintz (2000) and
Geffert (2002, priv. comm.) present B,V CCD observations for around
4450 stars (limited to V = 18.5 mag) covering a field of 20 x 20
arcminutes of M71.

\begin{figure}
\centering
\includegraphics[width=8cm]{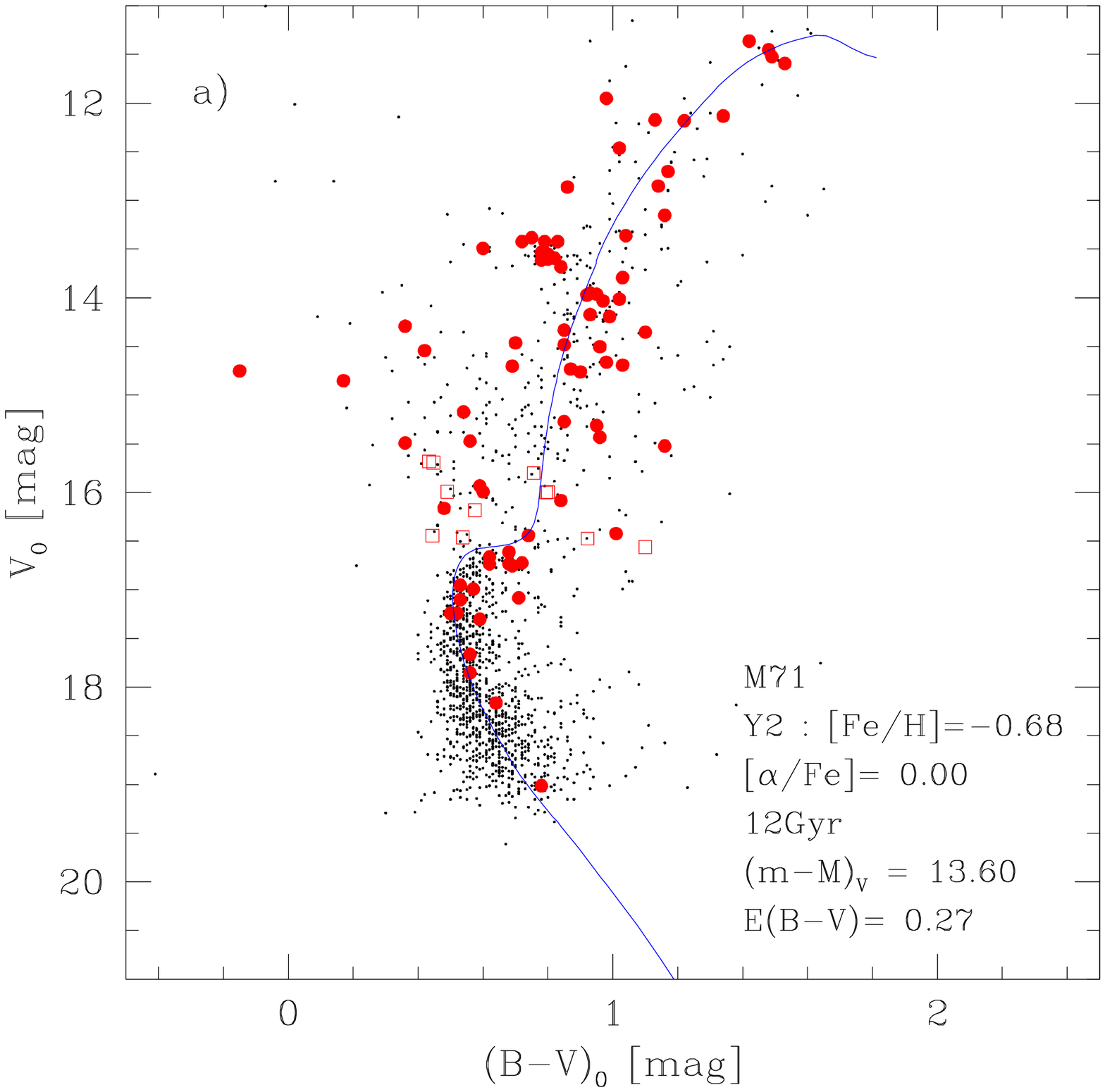}
\includegraphics[width=8cm]{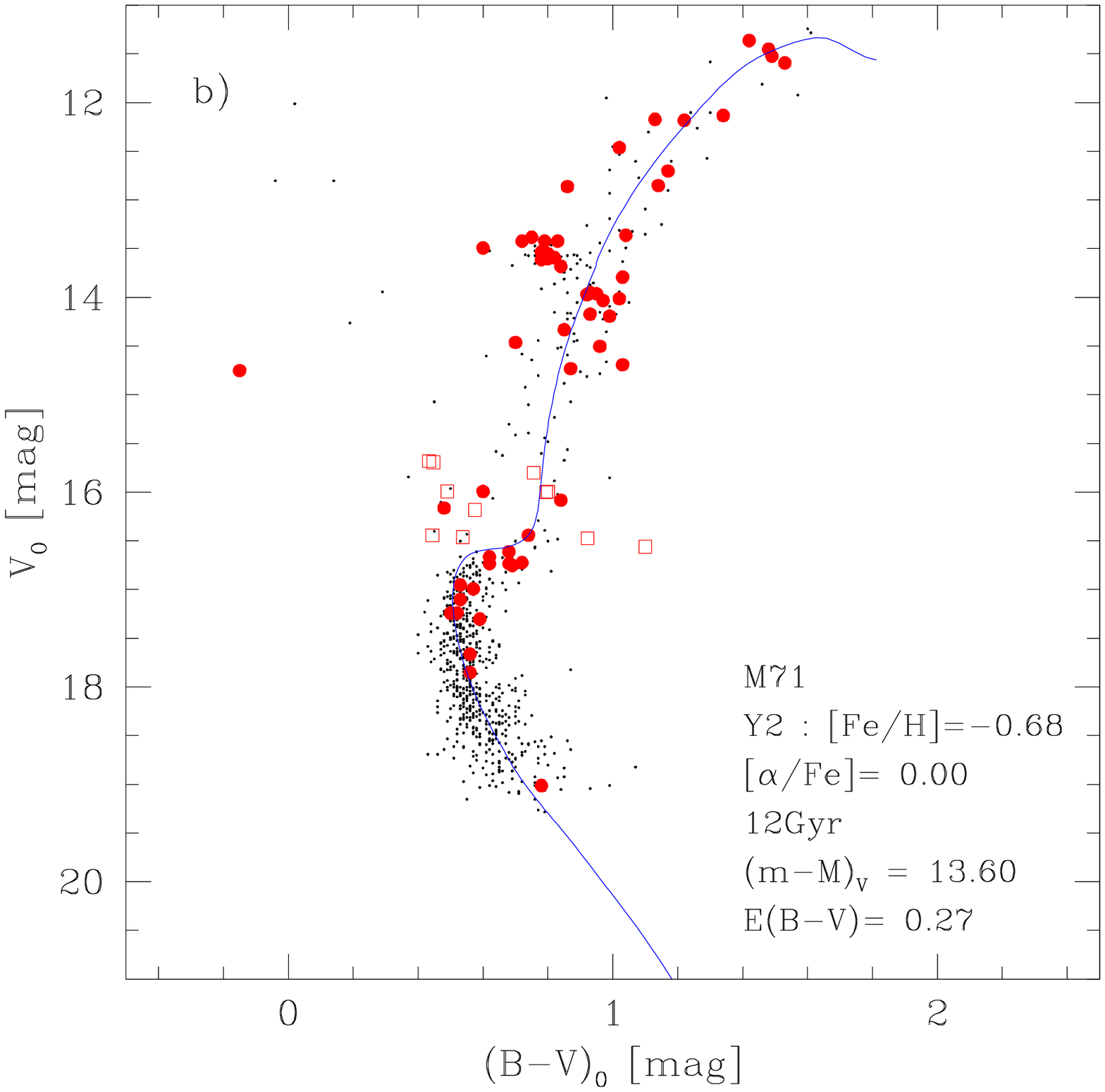}
\caption{The optical V,B-V CMD of M71. 
{\it Top}: {\it dots}:  Cudworth (1985,2006) photometry;
{\it filled circles}: sample stars; {\it open
squares}: stars identified in Geffert \& Maintz (2000).
{\it Bottom}:  stars 
 with membership probability P $\geq$ 80\% (symbols are the same as  above). 
A Yonsei-Yale isochrone by Kim et al. (2002) with parameters as indicated is
overplotted.}
\label{m71}
\end{figure}

The colour-magnitude diagram (CMD) of M71, using the photometry
from Cudworth (1985, 2006) and marking the sample stars is displayed
in Figure \ref{m71}.  Overplotted is a Yonsei-Yale isochrone (Kim
et al. 2002) with parameters [Fe/H] = $-$0.68 dex, [$\alpha $ /Fe]
= 0.00 dex, age = 12 Gyr.  The isochrone has been shifted by
(m-M)$_{\rm V}$ = 13.60 mag and E(B-V) = 0.27 mag (Geffert \& Maintz
2000).  The stars' designations, positions, membership
probabilities, colours, along with bolometric magnitudes, gravities
and temperatures are shown in Table \ref{parametros}.

\onllongtab{1}{
\begin{longtable}{lcccccccccccc}     
\caption{Program stars : star identification (1, as given in the observation
masks and Cudworth 1985;2006), coordinates J2000 (2)-(3), heliocentric radial
velocities [km/s] (4), membership probabilities (5, provided by Cudworth
1985;2006), V magnitude (6), colours(7)-(8), bolometric magnitudes (9),
gravities [dex](10), temperatures [K] (11).}
\label{parametros}   \\

\hline\hline    

\noalign{\smallskip}
\noalign{\vskip 0.1cm} 
ID & $\alpha$ & $\delta$ & v$_{r}$ & P & V & B-V & (B-V)$_{o}$ & Mbol & log g & T$_{\rm B-V}$ \\
\noalign{\vskip 0.1cm}
(1) & (2) & (3) & (4) & (5) & (6) & (7) & (8) & (9) & (10) & (11) \\
\noalign{\vskip 0.1cm}   

\hline 
\noalign{\vskip 0.1cm}
\endfirsthead

\multicolumn{11}{c}%
{{\tablename\ \thetable{} -- continued}} \\
\hline \hline 
ID & $\alpha$ & $\delta$ & v$_{r}$ & P & V & B-V & (B-V)$_{o}$ & Mbol & log g & T$_{\rm B-V}$ \\
\noalign{\vskip 0.1cm}
(1) & (2) & (3) & (4) & (5) & (6) & (7) & (8) & (9) & (10) & (11) \\
\noalign{\vskip 0.1cm} 
\hline \\
\endhead
\endfoot
\hline
\endlastfoot
\noalign{\vskip 0.1cm}
155	      : KC-141 &  298.4016  &  +18.75612  &	16.13  & 91	&  15.34 &  1.23  &  0.960  &  1.370 & 2.63 & 4709  \\
138  	      : KC-127 &  298.3995  &  +18.79042  &	-8.10  & 90	&  14.44 &  1.07  &  0.800  &  0.602 & 2.43 & 5000  \\
114  	      : KC-136 &  298.3962  &  +18.76293  &	16.62  & 93	&  14.80 &  1.22  &  0.950  &  0.839 & 2.43 & 4726  \\
253  	      : KC-119 &  298.4147  &  +18.81320  &    123.03  & 95	&  14.52 &  1.11  &  0.840  &  0.650 & 2.42 & 4924  \\
1468$^{\dag}$ : ...    &  298.4188  &  +18.77991  &	-4.63  & 80	&  17.30 &  0.81  &  0.540  &  3.723 & 3.94 & 5799  \\
367  	      : KC-343 &  298.4335  &  +18.79543  &	10.66  & 8	&  16.33 &  0.63  &  0.360  &  2.914 & 3.81 & 6500  \\
327  	      : 1-73   &  298.4268  &  +18.78301  &    -31.76  & 96	&  14.79 &  1.20  &  0.930  &  0.846 & 2.44 & 4761  \\
346  	      : KC-215 &  298.4298  &  +18.79586  &	37.87  & 93	&  15.30 &  0.97  &  0.700  &  1.539 & 2.88 & 5202  \\
356  	      : 1-71   &  298.4318  &  +18.79003  &    -22.45  & 96	&  13.54 &  1.44  &  1.170  & -0.627 & 1.71 & 4375  \\
221  	      : 1-109  &  298.4103  &  +18.77245  &	23.54  & 96	&  14.87 &  1.24  &  0.970  &  0.891 & 2.44 & 4691  \\
207  	      : 1639   &  298.4085  &  +18.79252  &	-4.87  & 95	&  17.00 &  0.75  &  0.480  &  3.477 & 3.90 & 6012  \\
291  	      : 1-95   &  298.4206  &  +18.76555  &	16.01  & 96	&  13.30 &  1.29  &  1.020  & -0.723 & 1.76 & 4608  \\
306  	      : 1-75   &  298.4229  &  +18.79264  &	58.09  & 96	&  14.85 &  1.29  &  1.020  &  0.827 & 2.38 & 4608  \\
751  	      : 1289   &  298.4129  &  +18.78219  &	-1.69  & 90	&  17.83 &  0.84  &  0.570  &  4.227 & 4.47 & 5705  \\
3791$^{\dag}$ : ...    &  298.4245  &  +18.79684  &	32.12  & 80	&  16.83 &  1.07  &  0.800  &  2.992 & 3.39 & 5000  \\
1936 	      : 1280   &  298.4169  &  +18.77425  &	 1.80  & 95	&  17.45 &  0.95  &  0.680  &  3.745 & 3.81 & 5366  \\
390  	      : 1-66   &  298.4382  &  +18.77940  &    -14.50  & 81	&  13.01 &  1.40  &  1.130  & -1.117 & 1.54 & 4435  \\
1511 	      : 1-113  &  298.4403  &  +18.79596  &    -11.98  & 95	&  12.43 &  1.80  &  1.530  & -2.222 & 0.87 & 3903  \\
1124 	      : 1-88   &  298.4363  &  +18.76084  &	 5.84  & 90	&  14.26 &  0.99  &  0.720  &  0.484 & 2.44 & 5161  \\
69   	      : KC-202 &  298.4608  &  +18.81061  &	43.91  & 94	&  15.17 &  1.12  &  0.850  &  1.292 & 2.67 & 4905  \\
7677 	      : 1346   &  298.4740  &  +18.79896  &	87.14  & 84	&  18.14 &  0.86  &  0.590  &  4.517 & 4.50 & 5634  \\
51$^{\dag}$   : ...    &  298.4469  &  +18.77311  &	-6.75  & 80	&  17.30 &  0.71  &  0.440  &  3.813 & 4.08 & 6165  \\
488$^{\dag}$  : ...    &  298.4532  &  +18.78136  &	-1.78  & 80	&  16.83 &  0.76  &  0.490  &  3.298 & 3.82 & 5976  \\
7277 	      : KC-336 &  298.4665  &  +18.79440  &	17.59  & ...	&  ...   &  ...   &  ...    & ... & ...  &	...   \\
1351 	      : 1-45   &  298.4507  &  +18.79796  &	39.18  & 94	&  12.36 &  1.76  &  1.490  & -2.221 & 0.89 & 3950  \\
573  	      : KC-196 &  298.4697  &  +18.79116  &	23.10  & 69	&  15.60 &  1.17  &  0.900  &  1.681 & 2.80 & 4814  \\
505  	      : 1-34   &  298.4574  &  +18.78078  &   -107.0   & 95	&  14.45 &  1.05  &  0.780  &  0.628 & 2.46 & 5039  \\
1556 	      : 1-46   &  298.4642  &  +18.79901  &	75.48  & 93	&  12.29 &  1.75  &  1.480  & -2.275 & 0.88 & 3962  \\
7498 	      : A4     &  298.4715  &  +18.77711  &	29.98  & 94	&  12.20 &  1.69  &  1.420  & -2.271 & 0.91 & 4036  \\
917  	      : KC-200 &  298.4624  &  +18.80419  &	62.65  & 86	&  15.53 &  1.30  &  1.030  &  1.498 & 2.64 & 4592  \\
1796 	      : 1134   &  298.4556  &  +18.80151  &	86.97  & 91	&  18.69 &  0.83  &  0.560  &  5.097 & 4.45 & 5741  \\
458  	      : 1-43   &  298.4485  &  +18.79111  &	27.59  & 94	&  14.26 &  1.06  &  0.790  &  0.430 & 2.37 & 5020  \\
6761 	      : ...    &  298.4591  &  +18.77784  &	31.48  & ...	&  ...   &  ...   &  ...    & ...  & ...  &	...   \\
566  	      : KC-363 &  298.4681  &  +18.77355  &    -74.33  & 0	&  16.36 &  1.43  &  1.160  &  2.203 & 2.85 & 4390  \\
439  	      : ...    &  298.4450  &  +18.77383  &	17.75  & ...	&  ...   &  ...   &  ...    & ... & ...  &	...   \\
619  	      : 1355   &  298.4770  &  +18.79714  &	12.77  & 97	&  17.94 &  0.80  &  0.530  &  4.377 & 4.37 & 5852  \\
1235 	      : 1670   &  298.4834  &  +18.79839  &	93.59  & 0	&  16.77 &  0.86  &  0.590  &  3.148 & 3.66 & 5635  \\
640  	      : 1-48   &  298.4814  &  +18.79968  &	62.93  & 94	&  14.39 &  1.07  &  0.800  &  0.552 & 2.41 & 5000  \\
122  	      : KC-303 &  298.3969  &  +18.77009  &    -30.64  & 0	&  16.01 &  0.81  &  0.540  &  2.433 & 3.42 & 5799  \\
1035 	      : 1600   &  298.3999  &  +18.77520  &    -17.97  & 97	&  17.57 &  0.95  &  0.680  &  3.855 & 4.59 & 5336  \\
787  	      : 1-59   &  298.4237  &  +18.80803  &	48.60  & 89	&  14.63 &  1.30  &  1.030  &  0.598 & 2.28 & 4592  \\
292  	      : X      &  298.4207  &  +18.73372  &    -46.00  & 90	&  14.40 &  1.07  &  0.800  &  0.562 & 2.42 & 5000  \\
258  	      : KC-152 &  298.4158  &  +18.73698  &	-2.72  & 85	&  15.03 &  1.26  &  0.990  &  1.034 & 2.48 & 4658  \\
2984$^{\dag}$ : ...    &  298.4104  &  +18.77345  &	-7.12  & 80	&  17.02 &  0.84  &  0.570  &  3.416 & 3.78 & 5699  \\
328$^{\dag}$  : ...    &  298.4270  &  +18.75018  &    -53.12  & 80	&  16.52 &  0.70  &  0.430  &  3.042 & 3.78 & 6205  \\
1060 	      : ...    &  298.4140  &  +18.77067  &	-2.37  & 80	&  ...   &  ...   &  ...    & ... & ... &      ...  \\
1294 	      : ...    &  298.4180  &  +18.78680  &	-5.85  & ...	&  ...   &  ...   &  ...    & ... & ... &      ...  \\
792$^{\dag}$  : ...    &  298.4254  &  +18.79479  &    -24.60  & 80	&  16.64 &  1.03  &  0.760  &  2.833 & 3.35 & 5079  \\
13   	      : 1-79   &  298.4087  &  +18.77783  &	34.37  &  51	&  13.99 &  1.43  &  1.160  & -0.167 & 1.90 & 4390  \\
232$^{\dag}$  : ...    &  298.4123  &  +18.77188  &    -11.50  & 80	&  16.84 &  1.06  &  0.790  &  3.010 & 3.40 & 5020  \\
402  	      : 1-87   &  298.4396  &  +18.76128  &    -12.73  & 95	&  14.37 &  1.05  &  0.780  &  0.548 & 2.42 & 5039  \\
4126 	      : ...    &  298.4290  &  +18.77043  &    -21.62  & ...	&  ...   &  ...   &  ...    & ... & ... &	...  \\
1633 	      : KC-374 &  298.4338  &  +18.78996  &	 2.45  & 0	&  16.27 &  1.23  &  0.960  &  2.300 & 3.01 & 4709  \\
2001 	      : 1-103  &  298.4317  &  +18.76827  &    -20.97  & 89	&  14.33 &  0.87  &  0.600  &  0.698 & 2.67 & 5604  \\
1873 	      : ...    &  298.4356  &  +18.78181  &    -31.61  & ...	&  ...   &  ...   &  ...    & ... & ... &	...  \\
10387	      : KC-267 &  298.4380  &  +18.77221  &	-34.8  & 39	&  15.69 &  0.44  &  0.170  &  2.443 & 3.87 & 7499  \\
1958 	      : ...    &  298.4417  &  +18.77365  &    -24.38  & ...	&  ...   &  ...   &  ...    & ... & ... &	...  \\
921$^{\dag}$  : ...    &  298.4650  &  +18.77283  &	-4.46  & 80	&  16.53 &  0.72  &  0.450  &  3.034 & 3.76 & 6126  \\
526  	      : 1-53   &  298.4607  &  +18.81613  &	40.67  & 95	&  12.97 &  1.61  &  1.340  & -1.392 & 1.31 & 4138  \\
484  	      : 1-55   &  298.4526  &  +18.80644  &	48.75  & 92	&  14.26 &  1.10  &  0.830  &  0.398 & 2.33 & 4943  \\
6548 	      : 2071   &  298.4558  &  +18.80534  &	-5.19  & 90	&  19.85 &  1.05  &  0.780  &  6.027 & 4.65 & 5039  \\
1181 	      : 1058   &  298.4628  &  +18.79180  &    -22.38  & 93	&  17.79 &  0.80  &  0.530  &  4.227 & 4.37 & 5852  \\
1785 	      : 1-36   &  298.4463  &  +18.77901  &    -28.97  & 64	&  12.79 &  1.25  &  0.980  & -1.197 & 1.60 & 4675  \\
11860	      : ...    &  298.4575  &  +18.80072  &	30.41  & ...	&  ...   &  ...   &  ...    & ... & ... &    ...  \\
5871 	      : ...    &  298.4479  &  +18.79313  &    -57.30  & ...	&  ...   &  ...   &  ...    & ... & ... &    ...  \\
6079 	      : ...    &  298.4502  &  +18.77050  &    -55.70  & ...	&  ...   &  ...   &  ...    & ... & ... &    ...  \\
7299 	      : ...    &  298.4678  &  +18.80609  &    -63.31  & ...	&  ...   &  ...   &  ...    & ... & ... &    ...  \\
7453 	      : ...    &  298.4694  &  +18.80521  &	36.21  & ...	&  ...   &  ...   &  ...    & ... & ... &    ...  \\
1214 	      : 2084   &  298.4746  &  +18.80516  &	32.43  & 60	&  19.00 &  0.91  &  0.640  &  5.327 & 4.56 & 5465  \\
951  	      : 2080   &  298.4727  &  +18.80355  &	48.11  & 82	&  18.08 &  0.77  &  0.500  &  4.546 & 4.31 & 5967  \\
2154 	      : ...    &  298.4710  &  +18.78277  &	-6.90  & ...	&  ...   &  ...   &  ...    & ... & ... &	...  \\
1219 	      : 1356   &  298.4765  &  +18.79601  &	61.95  & 0	&  17.92 &  0.98  &  0.710  &  4.173 & 4.62 & 5243  \\
1716 	      : ...    &  298.4827  &  +18.79184  &    -29.50  & ...	&  ...   &  ...   &  ...    & ... & ... &     ...  \\
8009 	      : ...    &  298.4805  &  +18.79990  &   -139.8   & ...	&  ...   &  ...   &  ...    & ... & ... &     ...  \\
654  	      : 1669   &  298.4843  &  +18.80037  &	52.82  & 0	&  17.26 &  1.28  &  1.010  &  3.246 & 3.35 & 4624  \\
1926 	      : 1633   &  298.4001  &  +18.78803  &    -22.54  & 97	&  17.28 &  1.01  &  0.740  &  3.489 & 3.63 & 5120  \\
3    	      : 1995   &  298.3983  &  +18.77371  &	 3.43  & 98	&  16.92 &  1.11  &  0.840  &  3.050 & 3.38 & 4924  \\
338  	      : 1-107  &  298.4283  &  +18.77441  &	 0.02  & 90	&  13.70 &  1.13  &  0.860  & -0.186 & 2.08 & 4887  \\
391  	      : KC-234 &  298.4385  &  +18.76578  &	18.32  & 81	&  15.59 &  0.12  &  0.150  &  ... & ... & ...  \\
223  	      : KC-300 &  298.4105  &  +18.77746  &	 5.90  & 26	&  16.11 &  1.12  &  0.850  &  2.232 & 3.05 & 4905  \\
314  	      : ...    &  298.4240  &  +18.79783  &    113.38  & ...	&  ...   &  ...   &  ...    & ... & ... &	...  \\
1325 	      : 1-70   &  298.4343  &  +18.78580  &	18.25  & 92	&  14.43 &  1.09  &  0.820  &  0.576 & 2.41 & 4962  \\
294  	      : 1308   &  298.4210  &  +18.79190  &	37.53  & 0	&  16.31 &  0.83  &  0.560  &  2.715 & 3.51 & 5732  \\
4448 	      : ...    &  298.4327  &  +18.78299  &	 0.11  & ...	&  ...   &  ...   &  ...    & ... & ... & ...  \\
10218	      : ...    &  298.4360  &  +18.78151  &    -10.03  & ...	&  ...   &  ...   &  ...    & ... & ... & ...  \\
3915 	      : ...    &  298.4260  &  +18.78303  &	29.04  & ...	&  ...   &  ...   &  ...    & ... & ... & ...  \\
813  	      : ...    &  298.4306  &  +18.77201  &	 5.21  & ...	&  ...   &  ...   &  ...    & ... & ... & ...  \\
266  	      : KC-223 &  298.4171  &  +18.77581  &	63.33  & 0	&  15.54 &  0.96  &  0.690  &  1.826 & 3.04 & 5338  \\
256  	      : KC-224 &  298.4154  &  +18.76982  &	61.72  & 0	&  15.13 &  0.63  &  0.360  &  1.714 & 3.33 & 6500  \\
1934 	      : ...    &  298.4136  &  +18.78399  &	 7.47  & ...	&  ...   &  ...   &  ...    & ... & ... &	...  \\
206  	      : 1604   &  298.4084  &  +18.77378  &	48.45  & 83	&  16.83 &  0.87  &  0.600  &  3.198 & 3.67 & 5604  \\
407  	      : 1-65   &  298.4402  &  +18.78101  &	 4.68  & 80	&  14.20 &  1.31  &  1.040  &  0.159 & 2.10 & 4575  \\
1693 	      : ...    &  298.4683  &  +18.80594  &	51.17  & ...	&  ...   &  ...   &  ...    & ... & ... &    ...  \\
1818 	      : ...    &  298.4662  &  +18.80561  &	81.33  & ...	&  ...   &  ...   &  ...    & ... & ... &    ...  \\
6868 	      : ...    &  298.4617  &  +18.80363  &	84.90  & ...	&  ...   &  ...   &  ...    & ... & ... &    ...  \\
1210 	      : 1342   &  298.4741  &  +18.80269  &	72.68  & 92	&  18.50 &  0.83  &  0.560  &  4.907 & 4.45 & 5741  \\
2013 	      : ...    &  298.4565  &  +18.80247  &    120.85  & ...	&  ...   &  ...   &  ...    & ... & ... &     ...  \\
1161 	      : ...    &  298.4523  &  +18.80226  &	72.23  & ...	&  ...   &  ...   &  ...    & ... & ... &     ...  \\
6786 	      : ...    &  298.4601  &  +18.73407  &    -11.68  & ...	&  ...   &  ...   &  ...    & ... & ... &     ...  \\
12299	      : ...    &  298.4637  &  +18.78218  &	12.36  & ...	&  ...   &  ...   &  ...    & ... & ... &     ...  \\
888$^{\dag}$  : ...    &  298.4543  &  +18.78301  &	 3.27  & 80	&  17.39 &  1.37  &  1.100  &  3.293 & 3.32 & 4481  \\
7505 	      : ...    &  298.4702  &  +18.79745  &	60.95  & ...	&  ...   &  ...   &  ...    & ... & ... &      ...  \\
7573 	      : ...    &  298.4724  &  +18.79126  &	93.07  & ...	&  ...   &  ...   &  ...    & ... & ... &      ...  \\
452  	      : 1-21   &  298.4475  &  +18.76873  &	 4.58  & 95	&  13.02 &  1.49  &  1.220  & -1.200 & 1.45 & 4302  \\
10927	      : ...    &  298.4456  &  +18.77864  &    219.80  & ...	&  ...   &  ...   &  ...    & ... & ... &     ...  \\
5964 	      : ...    &  298.4495  &  +18.77822  &	11.88  & ...	&  ...   &  ...   &  ...    & ... & ... &     ...  \\
1897 	      : ...    &  298.4584  &  +18.77371  &	-2.35  & ...	&  ...   &  ...   &  ...    & ... & ... &     ...  \\
613  	      : 1378   &  298.4761  &  +18.78647  &	27.41  & 97	&  17.57 &  0.89  &  0.620  &  3.917 & 4.53 & 5531  \\
641  	      : KC-191 &  298.4816  &  +18.79488  &	25.35  & 85	&  15.01 &  1.20  &  0.930  &  1.066 & 2.53 & 4761  \\
651  	      : KC-170 &  298.4838  &  +18.75471  &    -15.25  & 92	&  15.57 &  1.14  &  0.870  &  1.676 & 2.82 & 4868  \\
720  	      : KC-125 &  298.4006  &  +18.79778  &	57.44  & 38	&  15.32 &  1.12  &  0.850  &  1.442 & 2.73 & 4905  \\
4    	      : 2021   &  298.3983  &  +18.78664  &    -31.06  & 99	&  17.59 &  0.96  &  0.690  &  3.865 & 4.60 & 5305  \\
399  	      : 1-81   &  298.4392  &  +18.77778  &	-0.78  & 95	&  13.69 &  1.41  &  1.140  & -0.447 & 1.80 & 4419  \\
1282 	      : KC-298 &  298.4108  &  +18.79258  &    -20.66  & 8	&  16.15 &  1.22  &  0.950  &  2.189 & 2.97 & 4726  \\
1081 	      : ...    &  298.4221  &  +18.77992  &	-2.77  & ...	&  ...   &  ...   &  ...    & ... & ... &	...  \\
829  	      : KC-265 &  298.4352  &  +18.77724  &	-7.32  & 70	&  15.50 &  1.25  &  0.980  &  1.513 & 2.68 & 4675  \\
4482 	      : ...    &  298.4329  &  +18.79019  &    -21.82  & ...	&  ...   &  ...   &  ...    & ... & ...&   ...  \\
1486 	      : ...    &  298.4293  &  +18.77786  &    -16.28  & ...	&  ...   &  ...   &  ...    & ... & ...&   ...  \\
3909 	      : ...    &  298.4273  &  +18.78753  &    -17.13  & ...	&  ...   &  ...   &  ...    & ... & ...&   ...  \\
1087 	      : ...    &  298.4247  &  +18.77539  &	-7.65  & ...	&  ...   &  ...   &  ...    & ... & ...&   ...  \\
387  	      : ...    &  298.4374  &  +18.79451  &	-9.97  & ...	&  ...   &  ...   &  ...    & ... & ...&   ...  \\
778  	      : ...    &  298.4201  &  +18.78474  &    -22.07  & ...	&  ...   &  ...   &  ...    & ... & ...&   ...  \\
2045 	      : ...    &  298.4178  &  +18.78361  &    -26.96  & ...	&  ...   &  ...   &  ...    & ... & ...&   ...  \\
1292$^{\dag}$ : ...    &  298.4159  &  +18.77535  &    -10.03  & 80	&  17.31 &  1.19  &  0.920  &  3.374 & 3.46 & 4778  \\
242  	      : 1281   &  298.4135  &  +18.77435  &	21.63  & 94	&  17.56 &  0.99  &  0.720  &  3.803 & 4.62 & 5213  \\
212  	      : KC-128 &  298.4089  &  +18.78764  &    -21.42  & 0	&  15.19 &  1.37  &  1.100  &  1.093 & 2.44 & 4481  \\
5412 	      : ...    &  298.4417  &  +18.78268  &    -11.30  & ...	&  ...   &  ...   &  ...    & ... & ... &     ...  \\
574  	      : ...    &  298.4698  &  +18.80458  &	66.35  & ...	&  ...   &  ...   &  ...    & ... & ... &     ...  \\
78   	      : 1341   &  298.4731  &  +18.80265  &	51.13  & 96	&  18.08 &  0.79  &  0.520  &  4.526 & 4.34 & 5890  \\
1561 	      : ...    &  298.4658  &  +18.80221  &	55.75  & ...	&  ...   &  ...   &  ...    & ... & ... &    ...  \\
1894 	      : ...    &  298.4537  &  +18.80117  &	81.02  & ...	&  ...   &  ...   &  ...    & ... & ... &    ...  \\
881  	      : KC-252 &  298.4516  &  +18.78288  &    -19.77  & 0	&  15.38 &  0.69  &  0.420  &  1.911 & 3.34 & 6245  \\
7014 	      : ...    &  298.4625  &  +18.78173  &    -22.03  & ...	&  ...   &  ...   &  ...    & ... & ... & ...  \\
11907	      : ...    &  298.4568  &  +18.79274  &    -34.30  & ...	&  ...   &  ...   &  ...    & ... & ... & ...  \\
523  	      : ...    &  298.4605  &  +18.78475  &    -11.23  & ...	&  ...   &  ...   &  ...    & ... & ... & ...  \\
2229 	      : ...    &  298.4750  &  +18.78421  &	14.80  & ...	&  ...   &  ...   &  ...    & ... & ... & ...  \\
2153 	      : ...    &  298.4674  &  +18.79044  &	 1.29  & ...	&  ...   &  ...   &  ...    & ... & ... & ...  \\
513  	      : 1-11   &  298.4587  &  +18.76193  &	 1.72  & 87	&  14.81 &  1.19  &  0.920  &  0.874 & 2.46 & 4778  \\
5843 	      : 1-42   &  298.4474  &  +18.78542  &   -236.7   & 94	&  14.22 &  1.02  &  0.750  &  0.421 & 2.39 & 5099  \\
11331	      : ...    &  298.4495  &  +18.77523  &    -18.35  & ...	&  ...   &  ...   &  ...    & ... & ... &    ...  \\
1144 	      : ...    &  298.4455  &  +18.78746  &	-6.37  & ...	&  ...   &  ...   &  ...    & ... & ... &    ... \\
1815 	      : ...    &  298.4642  &  +18.77378  &    -12.25  & ...	&  ...   &  ...   &  ...    & ... & ... &    ... \\
616  	      : 1158   &  298.4767  &  +18.77737  &	-4.03  & 99	&  17.50 &  0.89  &  0.620  &  3.850 & 3.91 & 5542  \\
8076 	      : ...    &  298.4831  &  +18.79152  &	-8.45  & ...	&  ...   &  ...   &  ...    & ... & ... & ...	    \\
1409 	      : ...    &  298.4810  &  +18.79049  &	-5.78  & ...	&  ...   &  ...   &  ...    & ... & ... & ...	    \\

\hline     
\end{longtable} 
}

\subsubsection{Temperatures}

Effective temperatures (T$_{\rm eff}$) for the sample were obtained by
employing the empirical calibrations of Alonso et al. (1996; 1999; 2001) for
dwarfs and giants stars.
The intrinsic colours (B-V)$_{\rm o}$ were determined by adopting E(B-V) = 0.27
(Geffert \& Maintz 2000) and mean [Fe/H] = $-$0.73 (Harris 1996). 
The uncertainty in the effective temperatures  T(B-V) derived using Alonso et al.'s
calibrations is typically 150 K (within 1$\sigma$).

\subsubsection{Surface gravities}

 Photometric gravities for all stars were obtained in the
classical way, using the T$_{\rm eff}$s described in the previous
section and adopting $\rm M_* = 0.80~ M_{\odot}$, $\rm (m-M)_{\rm
V}$ = 13.60, E(B-V) = 0.27 (Geffert \& Maintz 2000), with bolometric
corrections taken from Alonso et al. (1999).  Input solar values
adopted were: $\rm T_{\odot} = 5780$ K, $\rm M_{\rm bol \odot}
= 4.75$, and $\log g$ = 4.44.  The error in $\rm M_{\rm bol}$ is
mostly due to the $\rm M_{\rm V}$ value and total extinction $\rm
A_{\rm V}$, the latter with an uncertainty around $\pm$0.05 mag.
For $\rm M_{\rm V}$, adopting an error in distance as large as 30\%,
we get $\rm \sigma_{M_{\rm bol}}$ $\sim $ 0.30 mag, which leads to
an error on the adopted photometric gravities of $\pm$0.30 dex.
 For stars fainter than V = 17.5, the surface gravities were
derived from  Yonsei-Yale 12 Gyr isochrones of Z=0.0040, 
Y = 0.24 and [Fe/H] = $-$0.70 (Kim et al. 2002).

\subsubsection{Spectral indices} 

The ability to analyze individual spectra of cluster members
can provide knowledge on the spectral properties of stellar
populations, placing constraints on stellar population synthesis.
In  medium-resolution spectra the measurement and analysis of
spectral indices is widely employed to interpret the chemical
evolution history of stellar populations in galaxies.

Absorption line indices were measured using the LECTOR program,
by A. Vazdekis, which measures line strengths in  1-D spectra
(Vazdekis et al. 2003).  Besides measuring the Lick/IDS
absorption-line indices (Worthey et al. 1994; Worthey \& Ottaviani
1997) we modified LECTOR in order to measure the line indices
recently defined by Serven et al. (2005).  In this paper, the only
index from the latter list we will discuss is Al3953, which was
shown by Serven et al.\ to be very sensitive to the abundance of
aluminium.

 For the uncertainties on the indices we initially considered
those provided by LECTOR (see Cardiel et al. 1998 for details).
These Poisson-based uncertainties on the indices are underestimated.
Thus, the uncertainties were also estimated from the standard
deviation of the measurements for each index on the individual
spectra, and hereafter these latest uncertainties will preferentially
be shown.

\begin{figure}
\centering
\includegraphics[width=8cm]{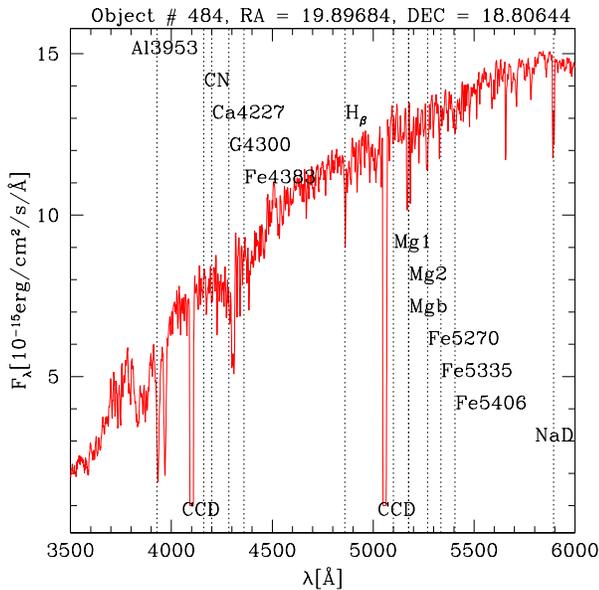}
\caption{Spectrum of the star 484 (1-55, V = 14.26) observed in the Mask \# 2.
Gaps are due to the 3 CCD configuration of the detector.
Central wavelengths of studied indices are indicated.}
\label{linhas}
\end{figure}

In Table \ref{ind} are presented the index passband definitions
where wavelengths are given in angstroms.  Tables \ref{tabind1},
\ref{tabind2} and \ref{tabind3} show the results and Poisson
uncertainties in the total-passband net counts for the indices
CN$_{1}$ [mag], CN$_{2}$ [mag], Ca4227 [$\rm \AA$], G4300:CH [mag],
Fe4383 [$\rm \AA$], H$_{\beta}$ [$\rm \AA$], Mg$_{1}$ [mag], Mg$_2$
[mag], Mgb [$\rm \AA$], Fe5270 [$\rm \AA$], Fe5335 [$\rm \AA$], Fe
5406 [$\rm \AA$], NaD [$\rm \AA$] and Al3953 [$\rm \AA$], while
Figure \ref{linhas} shows one of the final flux-calibrated spectra
of the sample with some features labeled.

\begin{table*}
\caption{Index definition.}             
\label{ind}      
\centering                          
\begin{tabular}{lccc}      
\hline\hline                
Index & Blue Pseudocontinuum & Feature & Red Pseudocontinuum \\   
\noalign{\vskip 0.1cm}
(1) & (2) & (3) & (4) \\ 
\hline                       
  
CN$_{1}$	&      4081.375 4118.875 &   4143.375 4178.375 &   4245.375 4285.375   \\
CN$_{2}$	&      4085.125 4097.625 &   4143.375 4178.375 &   4245.375 4285.375   \\
Ca4227		&      4212.250 4221.000 &   4223.500 4236.000 &   4242.250 4252.250   \\
G4300		&      4267.625 4283.875 &   4282.625 4317.625 &   4320.125 4336.375   \\
Fe4383		&      4360.375 4371.625 &   4370.375 4421.625 &   4444.125 4456.625   \\
H$_{-}$beta	&      4827.875 4847.875 &   4847.875 4876.625 &   4876.625 4891.625   \\
Mg$_{1}$	&      4895.125 4957.625 &   5069.125 5134.125 &   5301.125 5366.125   \\
Mg$_2$		&      4895.125 4957.625 &   5154.125 5196.625 &   5301.125 5366.125   \\
Mgb		&      5142.625 5161.375 &   5160.125 5192.625 &   5191.375 5206.375   \\
Fe5270		&      5233.150 5248.150 &   5245.650 5285.650 &   5285.650 5318.150   \\
Fe5335		&      5304.625 5315.875 &   5312.125 5352.125 &   5353.375 5363.375   \\
Fe5406		&      5376.250 5387.500 &   5387.500 5415.000 &   5415.000 5425.000   \\
NaD		&      5862.375 5877.375 &   5878.625 5911.125 &   5923.875 5949.875   \\
Al3953		&      3937.600 3967.400 &   3921.300 3935.500 &   3969.500 3987.000   \\

\hline                                  
\end{tabular}
\end{table*}

\onllongtab{3}{
\begin{longtable}{lccccccccccc}     
\caption{Selected spectral indices and uncertainties}        
\label{tabind1}   \\

\hline\hline    

\noalign{\smallskip}
\noalign{\vskip 0.1cm} 
ID & CN$_{1}$ & $\sigma$CN$_{1}$ & CN$_{2}$ & $\sigma$CN$_{2}$ & Ca4227 & $\sigma$Ca4227 & G4300 &
$\sigma$G4300 & Fe4383 & $\sigma$Fe4383  \\
\noalign{\vskip 0.1cm}
(1) & (2) & (3) & (4) & (5) & (6) & (7) & (8) & (9) & (10) & (11) & \\
\noalign{\vskip 0.1cm}

\hline 
\noalign{\vskip 0.1cm}
\endfirsthead

\multicolumn{11}{c}%
{{\tablename\ \thetable{} -- continued}} \\
\hline \hline 
ID & CN$_{1}$ & $\sigma$CN$_{1}$ & CN$_{2}$ & $\sigma$CN$_{2}$ & Ca4227 & $\sigma$Ca4227 & G4300 &
$\sigma$G4300 & Fe4383 & $\sigma$Fe4383 \\
\noalign{\vskip 0.1cm}
(1) & (2) & (3) & (4) & (5) & (6) & (7) & (8) & (9) & (10) & (11) \\
\noalign{\vskip 0.1cm} 
\hline \\
\endhead
\endfoot
\hline
\endlastfoot
\noalign{\vskip 0.1cm}
155  &  ...       &  ...       &  ...       &  ...       &    1.40243 &   0.06233 &    6.52152 &   0.10159 & 	4.20713 &  0.14444  \\
138  &   0.04322  &   0.00209  &   0.07336  &   0.00244  &    0.57001 &   0.03885 &    5.92503 &   0.05891 &   ...      &    ...    \\
114  &   0.16148  &   0.00286  &   0.20422  &   0.00334  &    0.65743 &   0.05216 &    5.96941 &   0.07767 & 	3.61326 &  0.11121  \\
253  &   0.00931  &   0.00229  &   0.03234  &   0.00269  &    0.61062 &   0.04255 &    6.61777 &   0.06490 & 	2.60329 &  0.09691  \\
1468 &  -0.05119  &   0.00791  &  -0.02598  &   0.00911  &    1.20442 &   0.13622 &    3.78093 &   0.23611 & 	1.77363 &  0.32466  \\
367  &  -0.11786  &   0.00402  &  -0.07367  &   0.00460  &    0.22357 &   0.07454 &    9.08490 &   0.15210 & 	0.83578 &  0.18196  \\
327  &   0.15312  &   0.00276  &   0.19179  &   0.00323  &    0.83128 &   0.05027 &    6.08654 &   0.07403 & 	3.37153 &  0.10663  \\
346  &  -0.03214  &   0.00320  &  -0.00845  &   0.00374  &    1.02249 &   0.05736 &     ...    &     ...   & 	2.97693 &  0.13212  \\
\hline     
\end{longtable} 
}

\onllongtab{4}{
\begin{longtable}{lccccccccccc}     
\caption{Selected specral indices and uncertainties.}        
\label{tabind2}   \\

\hline\hline    

\noalign{\smallskip}
\noalign{\vskip 0.1cm} 
ID & H$_{\beta}$ & $\sigma$H$_{\beta}$ & Mg$_{1}$ & $\sigma$Mg$_{1}$ & Mg$_2$ & $\sigma$Mg$_2$ & Mgb & $\sigma$Mgb & Fe5270 &
$\sigma$Fe5270 \\
\noalign{\vskip 0.1cm}
(1) & (2) & (3) & (4) & (5) & (6) & (7) & (8) & (9) & (10) & (11) \\
\noalign{\vskip 0.1cm}   

\hline 
\noalign{\vskip 0.1cm}
\endfirsthead

\multicolumn{11}{c}%
{{\tablename\ \thetable{} -- continued}} \\
\hline \hline 
ID & H$_{\beta}$ & $\sigma$H$_{\beta}$ & Mg$_{1}$ & $\sigma$Mg$_{1}$ & Mg$_2$ & $\sigma$Mg$_2$ & Mgb & $\sigma$Mgb & Fe5270 &
$\sigma$Fe5270 \\
\noalign{\vskip 0.1cm}
(1) & (2) & (3) & (4) & (5) & (6) & (7) & (8) & (9) & (10) & (11) \\
\noalign{\vskip 0.1cm} 
\hline \\
\endhead
\endfoot
\hline
\endlastfoot
\noalign{\vskip 0.1cm}
155  &   1.23728 &  0.05199  &   0.08470  & 0.00103 &	0.12360  & 0.00118 &   2.68898 & 0.04557 &   2.22092 & 0.04909 \\
138  &   1.71328 &  0.03203  &  -0.12243  & 0.00070 &  -0.11571  & 0.00082 &   1.48913 & 0.02953 &   1.77792 & 0.03179 \\
114  &   1.24208 &  0.04006  &  -0.15352  & 0.00085 &  -0.00161  & 0.00100 &   2.54644 & 0.03525 &   2.30571 & 0.03787 \\
253  &   1.38706 &  0.03437  &  -0.20548  & 0.00073 &  -0.07328  & 0.00086 &   1.51620 & 0.03057 &   1.78977 & 0.03300 \\
1468 &   2.65934 &  0.11462  &  -0.00489  & 0.00239 &	0.05761  & 0.00277 &   1.93860 & 0.10763 &   1.13950 & 0.12007 \\
367  &   4.51381 &  0.06618  &  -0.01103  & 0.00146 &	0.03080  & 0.00170 &   0.83520 & 0.06797 &  ...      &    ...   \\
327  &   1.09457 &  0.03862  &   0.02683  & 0.00076 &	0.11273  & 0.00088 &   2.63517 & 0.03389 &   2.26211 & 0.03668 \\
346  &   1.56316 &  0.04784  &   0.01315  & 0.00095 &	0.09936  & 0.00110 &   2.33412 & 0.04280 &  ...      &    ...   \\
\hline     
\end{longtable} 
}

\onllongtab{5}{
\begin{longtable}{lcccccccc}     
\caption{Selected spectral indices and uncertainties.}        
\label{tabind3} \\

\hline\hline    

\noalign{\smallskip}
\noalign{\vskip 0.1cm} 
ID & Fe5335 & $\sigma$Fe5335 & Fe5406 & $\sigma$Fe5406 & NaD &
$\sigma$NaD & Al3953 & $\sigma$Al3953 \\
\noalign{\vskip 0.1cm}
(1) & (2) & (3) & (4) & (5) & (6) & (7) & (8) & (9) \\
\noalign{\vskip 0.1cm}   

\hline 
\noalign{\vskip 0.1cm}
\endfirsthead

\multicolumn{9}{c}%
{{\tablename\ \thetable{} -- continued}} \\
\hline \hline 
ID & Fe5335 & $\sigma$Fe5335 & Fe5406 & $\sigma$Fe5406 & NaD &
$\sigma$NaD & Al3953 & $\sigma$Al3953 \\
\noalign{\vskip 0.1cm}
(1) & (2) & (3) & (4) & (5) & (6) & (7) & (8) & (9) \\
\noalign{\vskip 0.1cm} 
\hline \\
\endhead
\endfoot
\hline
\endlastfoot
\noalign{\vskip 0.1cm}
155  &    1.88008 & 0.05544&  1.13882 & 0.04054 &   2.04970 &	0.03406 &  ...      &   ...      \\  
138  &    ...     &    ... &  0.81074 & 0.02593 &   4.12518 &	0.02094 &   4.13757 &    0.08959 \\
114  &    1.97289 & 0.04263&  1.02490 & 0.03133 &  ...      &    ...    &  ...      &   ...      \\
253  &    1.51201 & 0.03747&  0.69764 & 0.02730 &  ...      &  ...      &  ...      &    ...     \\
1468 &    0.84823 & 0.13686&  0.69834 & 0.09912 &   1.86168 &	0.08544 &   2.70808 &    0.43285 \\
367  &    0.60807 & 0.08541&  0.20029 & 0.06286 &   1.48435 &	0.05619 &   3.41298 &    0.14671 \\
327  &    1.95244 & 0.04120&  1.13027 & 0.03021 &   3.26561 &	0.02444 &   3.25623 &    0.14144 \\
346  &    1.06836 & 0.05323&  0.87886 & 0.03839 &   1.95114 &	0.03249 &   3.73815 &    0.15558 \\
\hline     
\end{longtable} 
}

\section{Results}

The final sample adopted 
consists of stars with a membership probability higher than 80\% and
magnitudes V $\leq$ 15.5 mag. 
In addition, each spectrum was individually inspected by the
region of the index passbands. 
Below we present the main results.

\subsection{CN and CH}

The Lick CN indices measure the strength of the CN 4150 $\rm
\AA$ bandhead. Although we have measured CN$_{1}$ and CN$_{2}$, we
chose to work with CN$_{1}$ more extensively. Both indices
are similar in definition with a slight difference in their blue
pseudocontinuum definition.

\begin{figure}
\centering
\includegraphics[width=8cm]{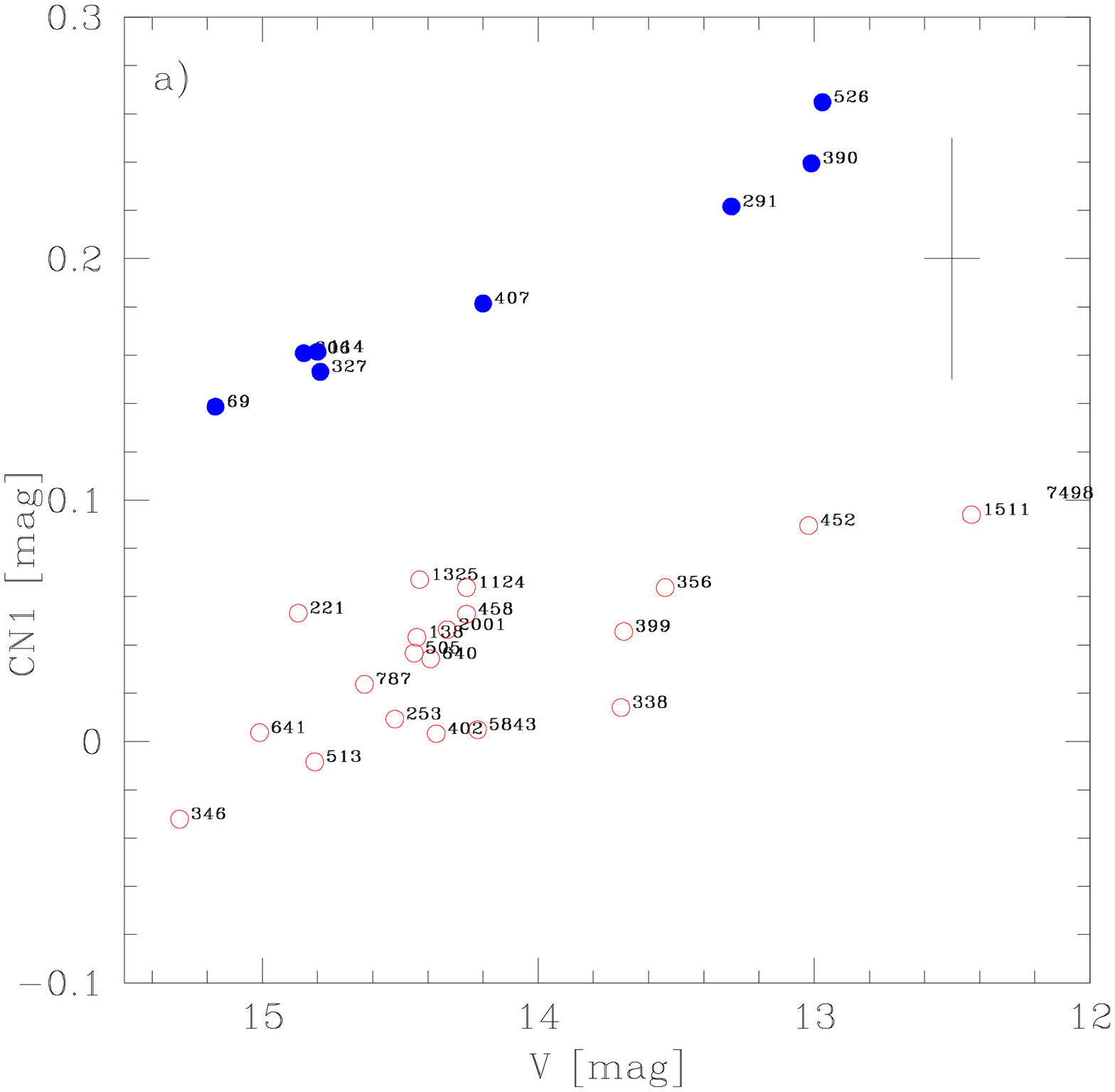}
\includegraphics[width=8cm]{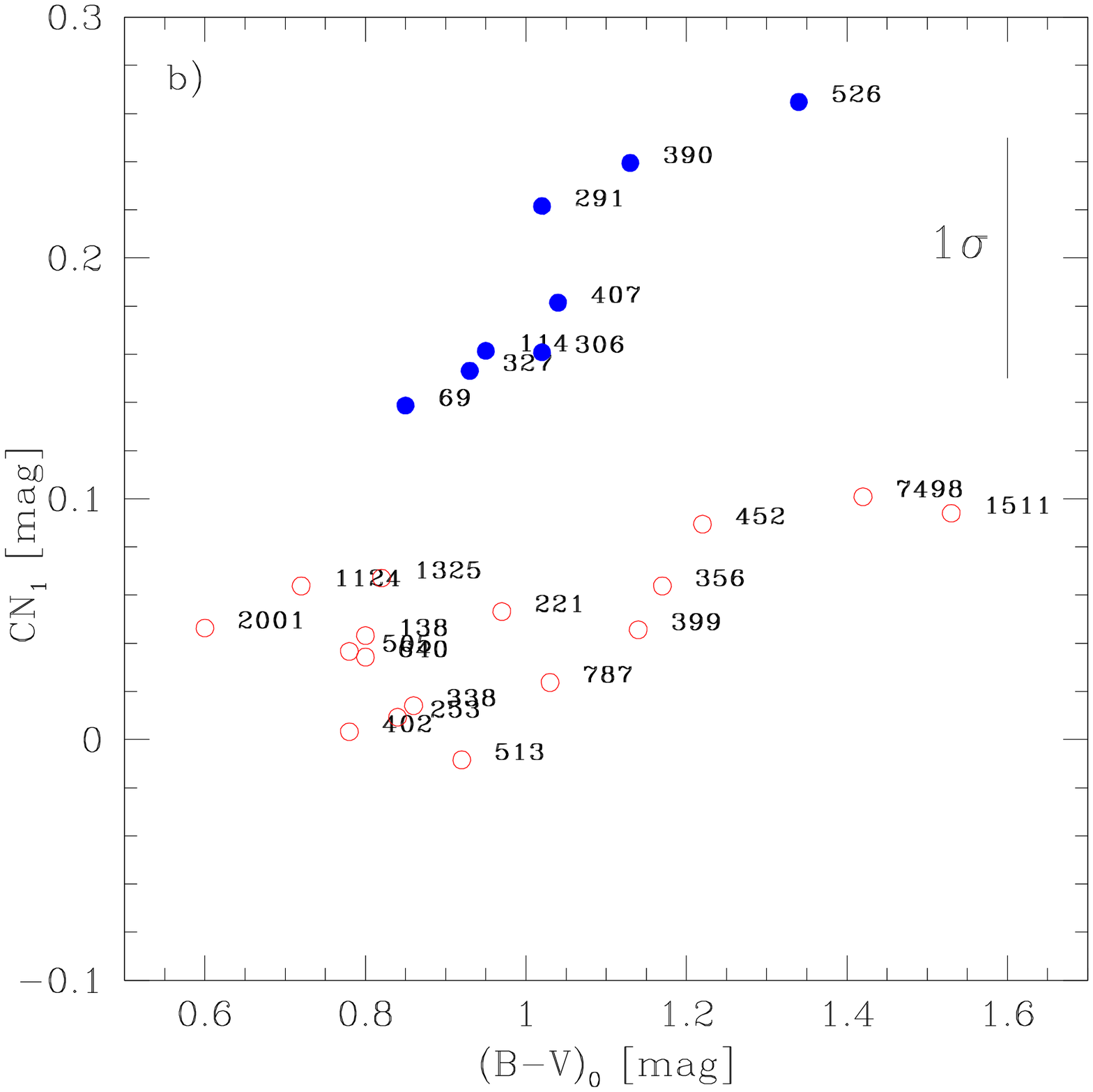}
\caption{{\it a)}: CN$_{1}$ plotted against V, and {\it b)}: CN$_{1}$ vs.
(B-V)$_{\rm 0}$ for the M71 sample stars, where the locus CN-strong ({\it filled
circles}) and CN-weak ({\it open circles}) are seen.
Error bars quoted correspond to the rms on all measurements available.}
\label{indcn}
\end{figure}

Figures \ref{indcn}a,b present the CN$_{1}$ index vs. V-band magnitude
and (B-V)$_{\rm 0}$ colours for M71 selected stars. Stellar
IDs are also marked.  The distribution of the data points in
these two plots is clearly bi-modal, with two families of stars
with strong and weak CN bands at fixed colour and/or magnitude.
The dividing line between CN-strong stars, it the top half, and
CN-weak stars in the bottom half of the plots runs diagonally from
the lower left to the top right of both plots.  This is because in
both families of stars, CN is a strong function of temperature, so
that CN-bands become stronger for lower temperatures in both the
CN-strong and CN-weak groups.
It is interesting that stars 291, 390 and 526
are AGB members, and they show a higher N enhancement.

Concerning stars 1351, 1556 and 640, their
CN measurements were slightly affected by the presence of the 2 CCD
gaps around their red passband. 
For the CN-strong stars, showing nitrogen excesses,
we found $\langle$CN$\rangle$ = 0.19 $\pm$ 0.04 (N = 8 stars),
whereas for the CN-weak stars $\langle$ CN $\rangle$ = 0.04 $\pm$
0.04 (N = 20 stars). Note that stars 1351, 1556 and 640 were not
taken into consideration in computing the mean CN-weak value.

\begin{figure}
\centering
\includegraphics[width=8cm]{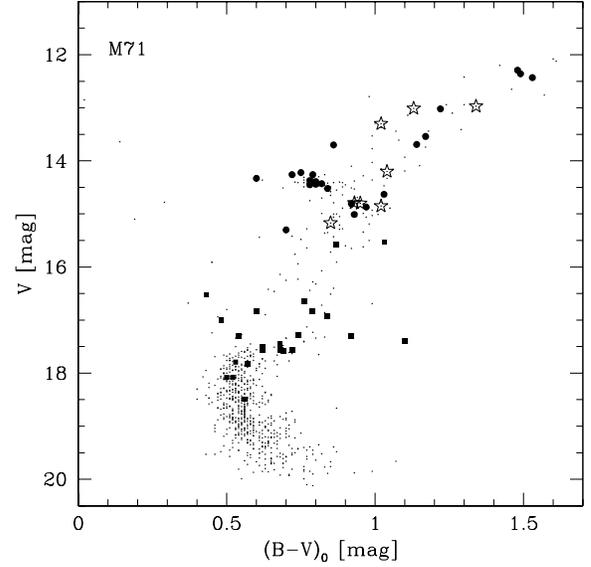}
\caption{V, B-V CMD with the CN-strong stars ({\it open stars}),
CN-weak ({\it filled circles}), and the subgiants ({\it \bf filled squares}) for which the CN measurements
are not qualified as strong or weak.}
\label{cmdcn}
\end{figure}

\begin{figure}
\centering
\includegraphics[width=8cm]{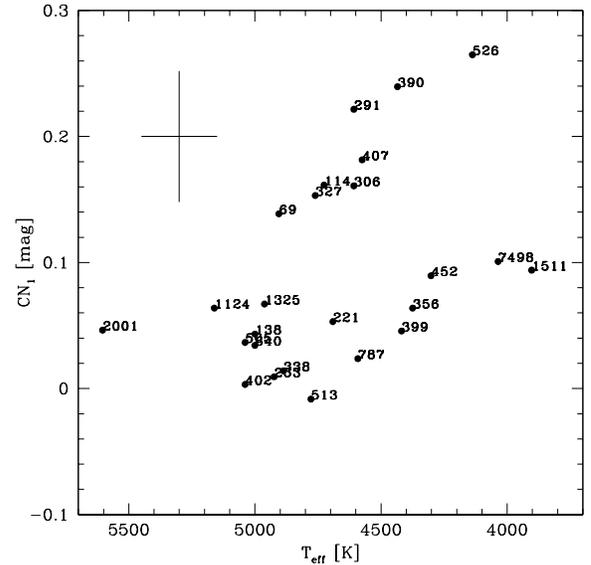}
\caption{CN$_{1}$ vs. effective temperature. Error bar on CN$_{1}$ is as explained 
in Fig. \ref{indcn}, while on T$_{\rm eff}$ it resembles the 
value determined with the photometric calibration employed.}
\label{indcntemp}
\end{figure}

\begin{figure}
\centering
\includegraphics[width=8cm]{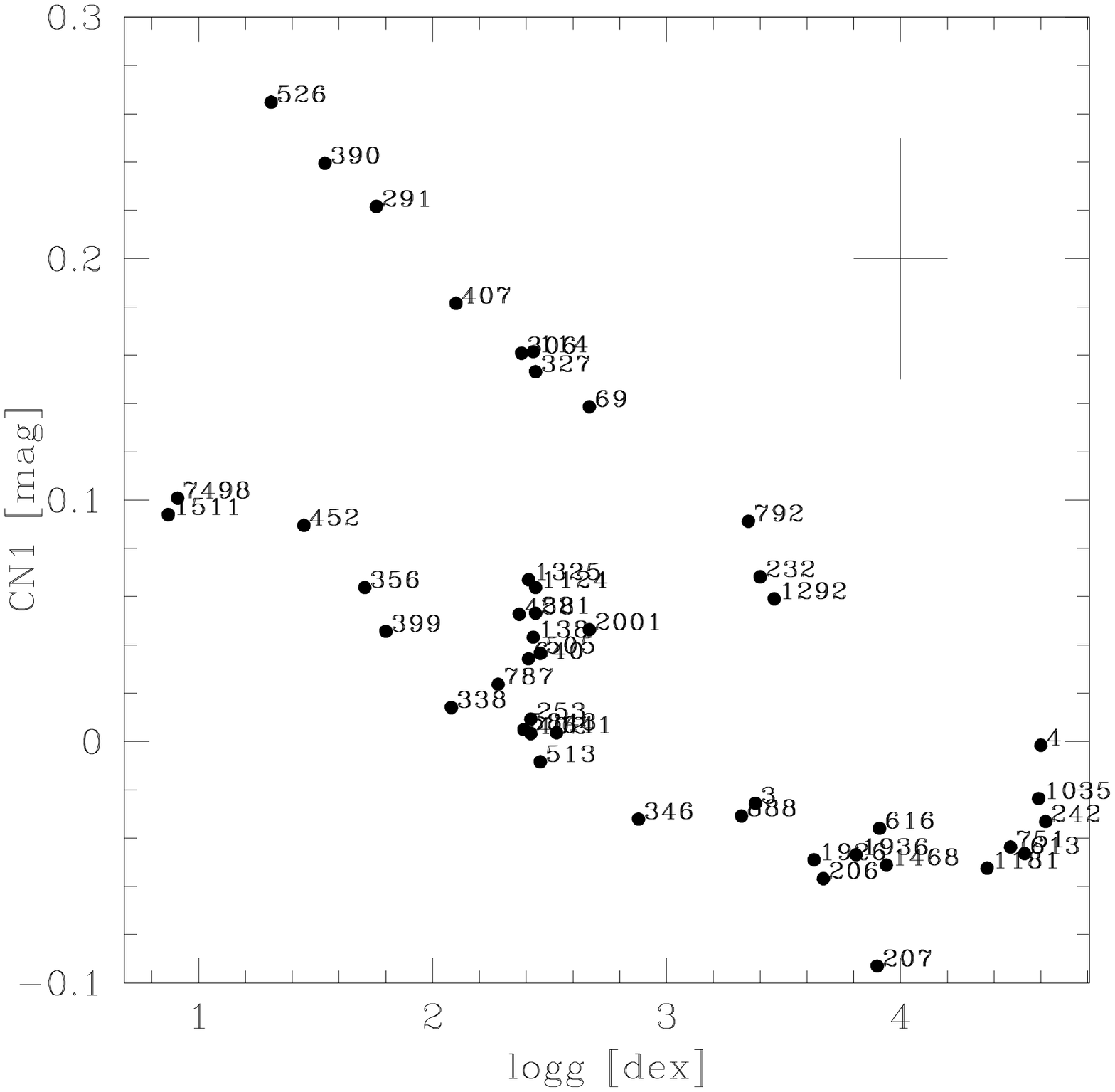}
\caption{CN$_{1}$ vs. log g. Error bar is indicated. }
\label{cn1logg}
\end{figure}

In Fig. \ref{cmdcn} we show the location of CN-strong and CN-weak
stars in the CMD, for V$<$15.5. Stars weaker than V$>$15.5 are also
plotted, with no distinction between CN intensities, given the lower
S/N of the spectra.
In Fig. \ref{indcntemp} the CN$_1$ index vs. T$_{\rm eff}$ is shown
for the giants, where the bimodality of CN-strength is found among
both RGB and Asymptotic Giant Branch (AGB) stars.
Fig. \ref{cn1logg} shows  CN$_1$ vs. log g for all stars, illustrating
the clear separation between CN-strong and CN-weak giants, whereas
for dwarfs, the bimodality is not  so clear, partly due to the
lower S/N of our dwarf-star spectra, and partly because the differences
between CN-strong and CN-weak spectra are more subtle, given that
CN-bands are weak overall in the spectra of warm turnoff stars.
Recall however that Cohen (1999) has found a bimodality for  main
sequence (MS) stars of M71.

\begin{figure}
\centering
\includegraphics[width=8cm]{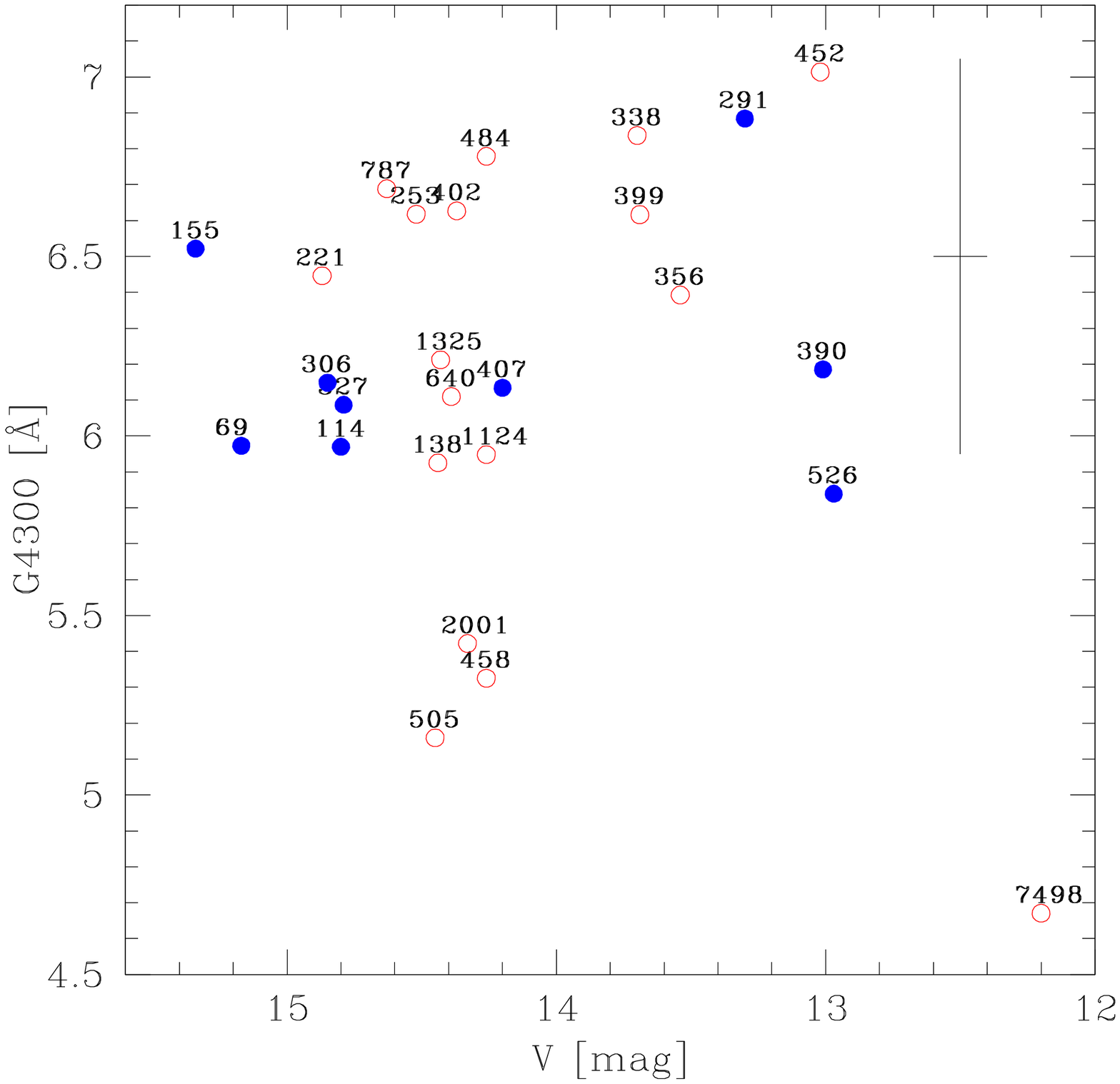}
\caption{G-band as function of V  magnitude, where CN-strong ({\it filled
circles}) and CN-weak ({\it open circles}) stars are plotted separately. }
\label{indch}
\end{figure}

\begin{figure}
\centering
\includegraphics[width=8cm]{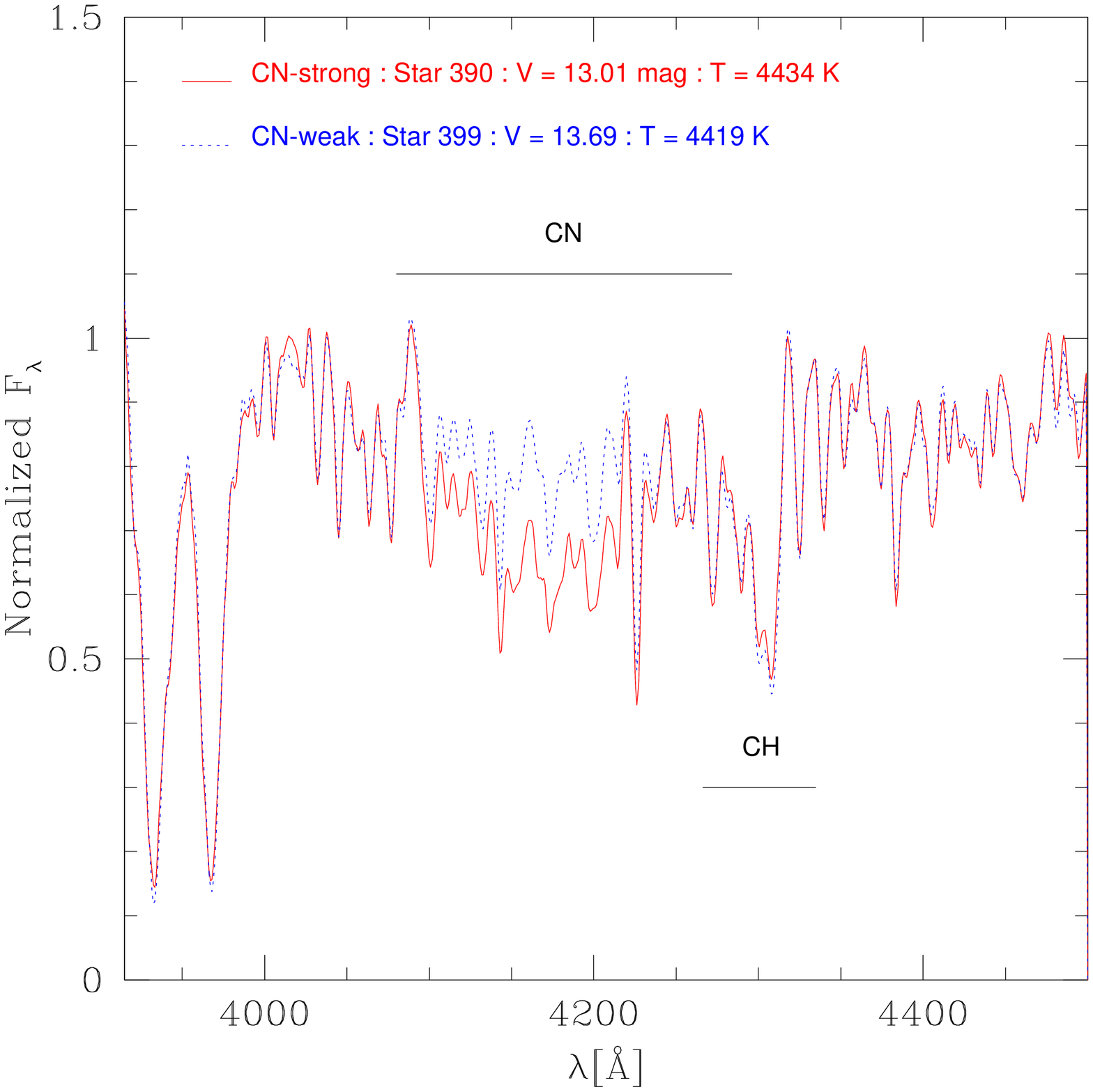}
\caption{Stars 390 and 399 in the molecular CN and CH band region.
The full and dashed lines represent the CN-strong and CN-weak stars,
respectively.  These stars display approximately the same temperatures.}
\label{cnchbands}
\end{figure}

The G-band at 4300 {\rm \AA} includes  a CH bandhead and can
be used to derive the carbon abundance.  Its behaviour along the main
sequence, subgiant and red giant sequences can give important clues
on mixing processes along stellar evolution.  In Fig. \ref{indch}
we plot the G band indices as a function of V magnitude. The symbols
are the same as used in Figs.  \ref{indcn}a,b, where the CN-strong and
CN-weak sequences are represented by filled and open
circles.  One can see that CN-strong stars tend to present lower
CH values, while the CN-weak stars appear to show higher values of
CH.  This general trend confirms the well-known CN-CH anticorrelation
in giant stars reported in past decades for many Galactic
globular clusters (e.g., Dickens et al. 1979; Smith 1987; Kraft
1994; Gratton et al. 2004 and references therein).  Note, however,
that despite their high membership probabilities, stars 2001, 458,
505 and 7498 present lower  G-band strengths than other CN-weak
stars of the sample. Stars 2001, 458 and 505 are HB members, and this suggests that
we also detect a bimodal distribution  in the HB. 
 Figure \ref{cnchbands} plots an example of CN-bimodality
and CN-CH anticorrelation for two M71 stars of similar V magnitude
and T$_{\rm eff}$. The CN-strong giant 390 (T = 4435 K : CN = 0.24
: CH = 6.18) presents a  stronger CN band and a weaker CH band
strength than star 399 (T = 4419 : CN = 0.04 : CH = 6.62).


\subsection{Iron and magnesium indices}

\begin{figure}
\centering
\includegraphics[width=8cm]{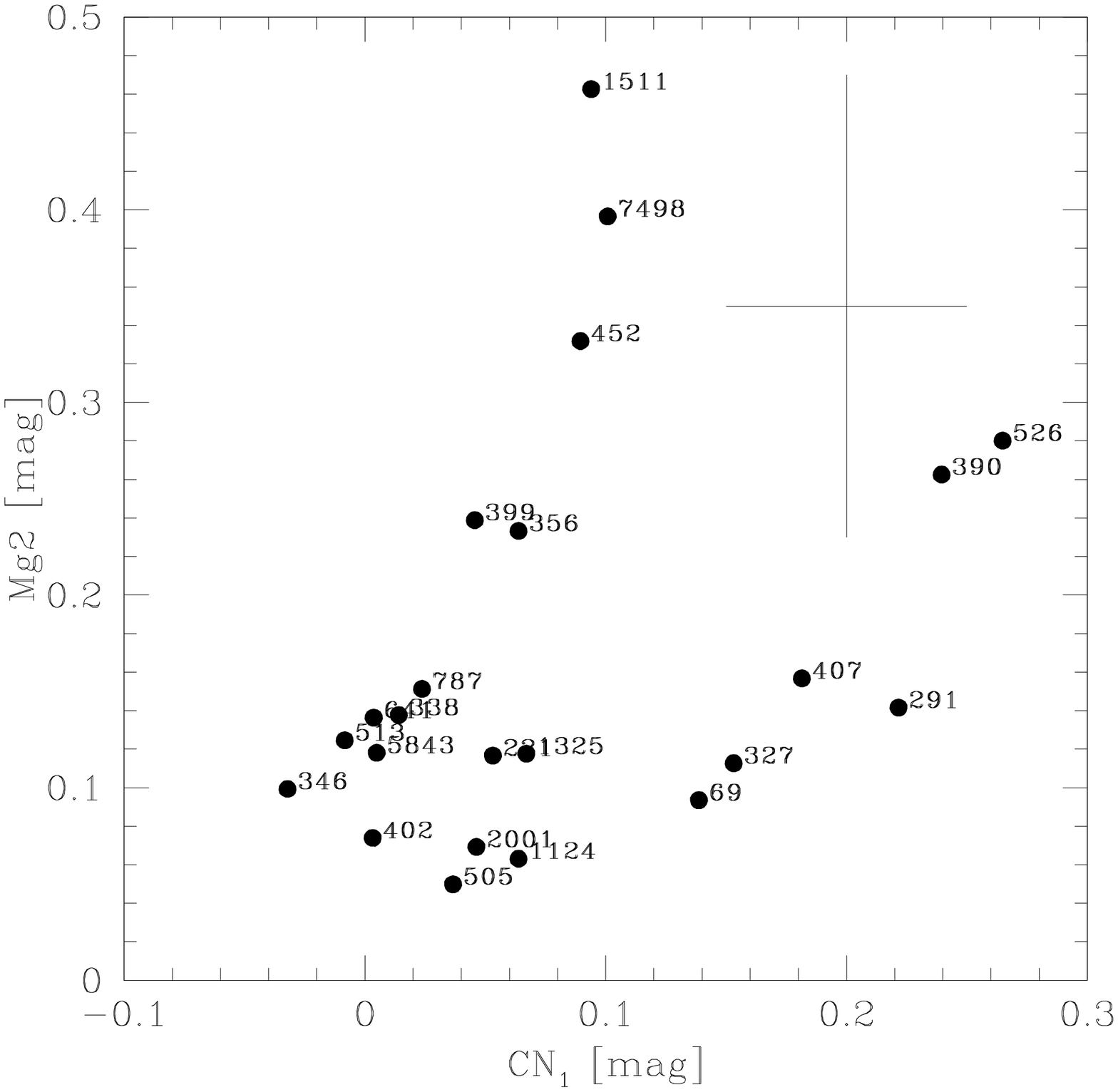}
\includegraphics[width=8cm]{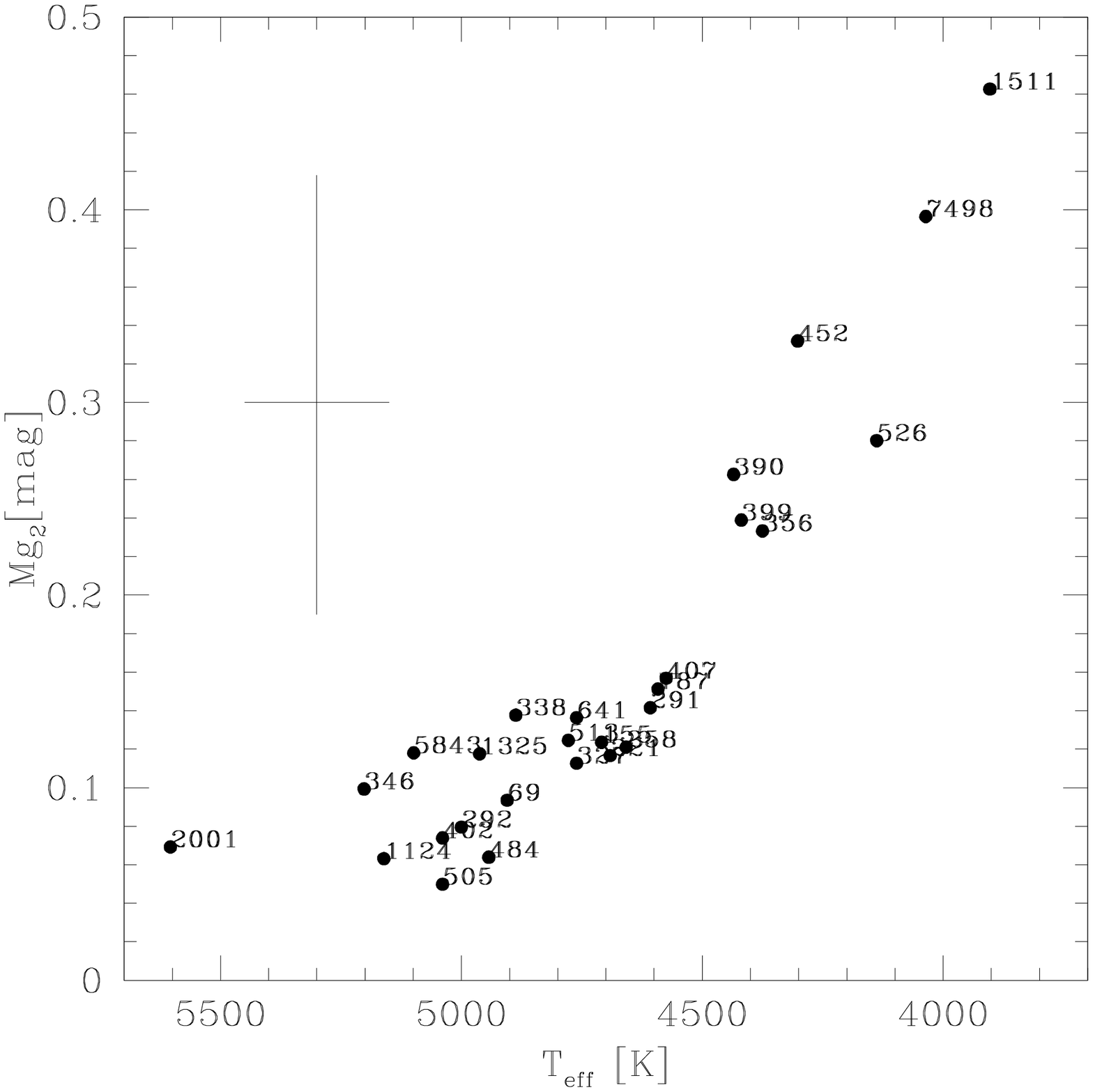}
\caption{{\it Upper panel}: Mg$_2$ plotted against CN$_{1}$;
{\it Lower panel}: Mg$_2$ as a 
function of the effective temperature. Error bars are as previously discussed.}
\label{indmg2cntemp}
\end{figure}

We measured spectral indices sensitive to the abundances of
iron (Fe4383, Fe5270, Fe5335, and Fe5406) and magnesium (Mg$_1$,
Mg$_2$ and Mgb) in all our spectra.  We henceforth focus on an
average Fe index, defined as $\langle$Fe$\rangle$ = (Fe4383 + Fe5270
+ Fe5335 + Fe5406)/4 and Mg$_2$. All these indices are of widespread
use in stellar population studies in galaxies  and stellar
clusters.  Figure \ref{indmg2cntemp} plots the behaviour of Mg$_2$
as a function of CN and T$_{\rm eff}$, respectively.  In this figure
we see that both CN-weak and CN-strong sequences correspond to
Mg$_2$ $\geq$ 0.05 mag and Mg$_2$ tends to be higher for cooler
stars (T$_{\rm eff}$ $\leq$ 4300 K).
Therefore, Mg$_2$ is weaker in CN-strong stars, and this seems
to indicate that besides a Mg-Al anticorrelation (Fig. \ref{indalnamg}),
and a weak Al-CN anticorrelation (Fig. \ref{indnacnal}),
we found some evidence for a  Mg-N anticorrelation
(see Gratton et al. 2004).

\subsection{Ca4227}

This index is widely used in stellar population studies in galaxies, where Ca
is considered as a representative $\alpha$-element (e.g. Thomas et al. 2003;
Prochaska et al. 2005). We found $\langle$Ca4227$\rangle$ = 1.25 $\pm$ 1.00 $\rm
\AA$ for 32 stars in the range 12 $\leq$ V $\leq$ 15.5 and with
a membership probability higher than 80\%.

\begin{figure}
\centering
\includegraphics[width=8cm]{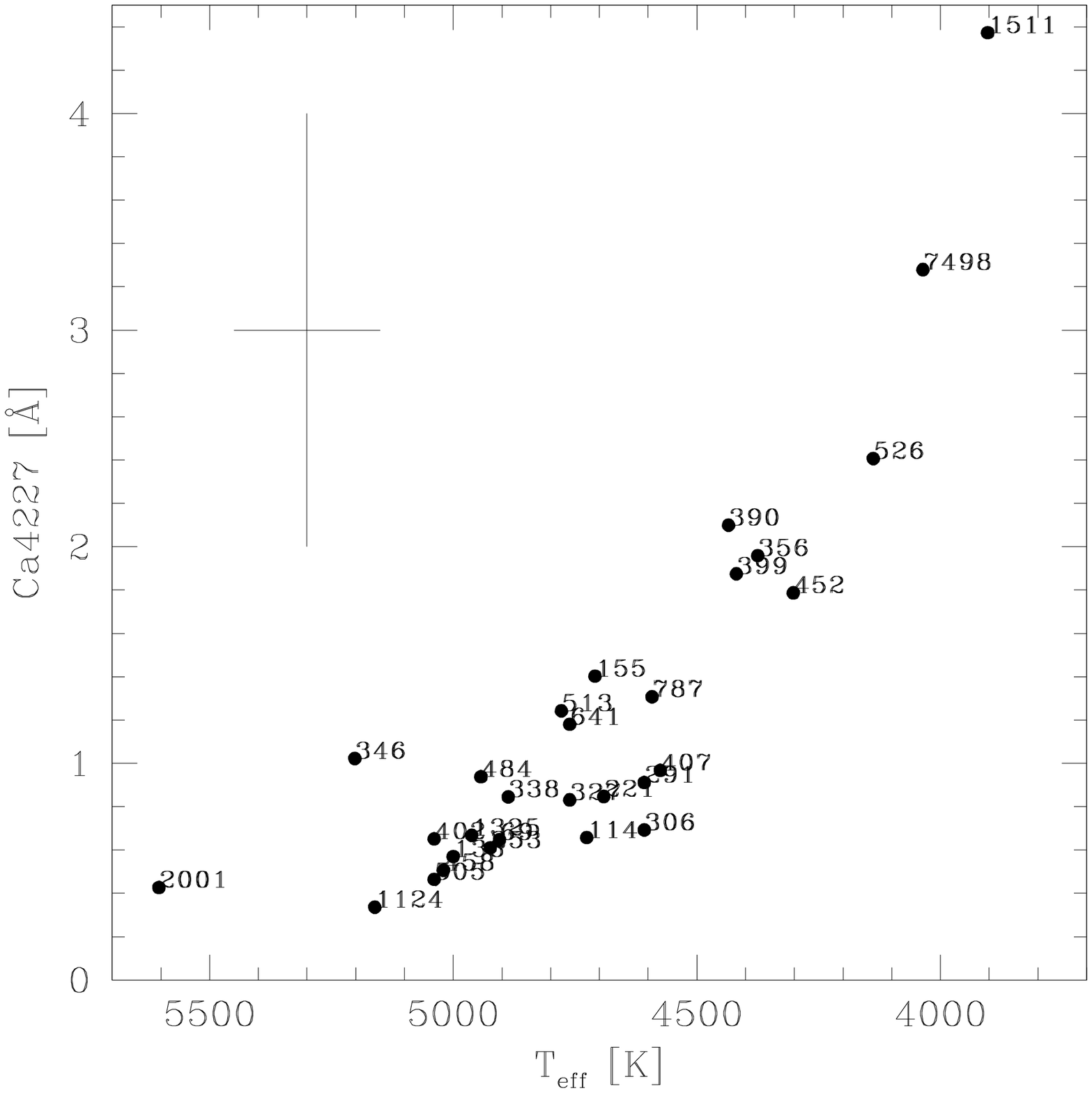}
\caption{Ca4227 feature against the effective temperature.}
\label{indcatemp}
\end{figure}

\begin{figure}
\centering
\includegraphics[width=8cm]{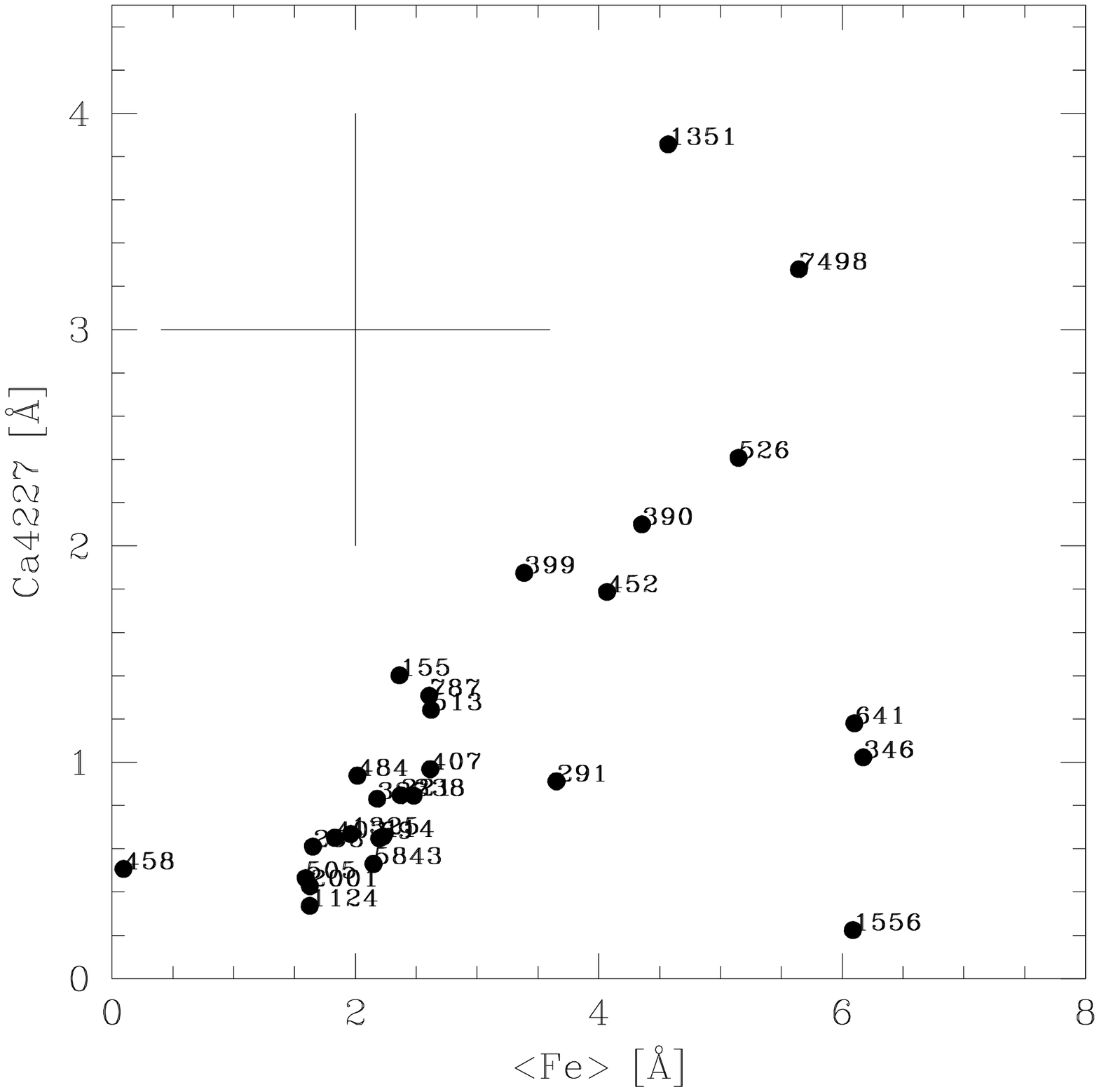}
\includegraphics[width=8cm]{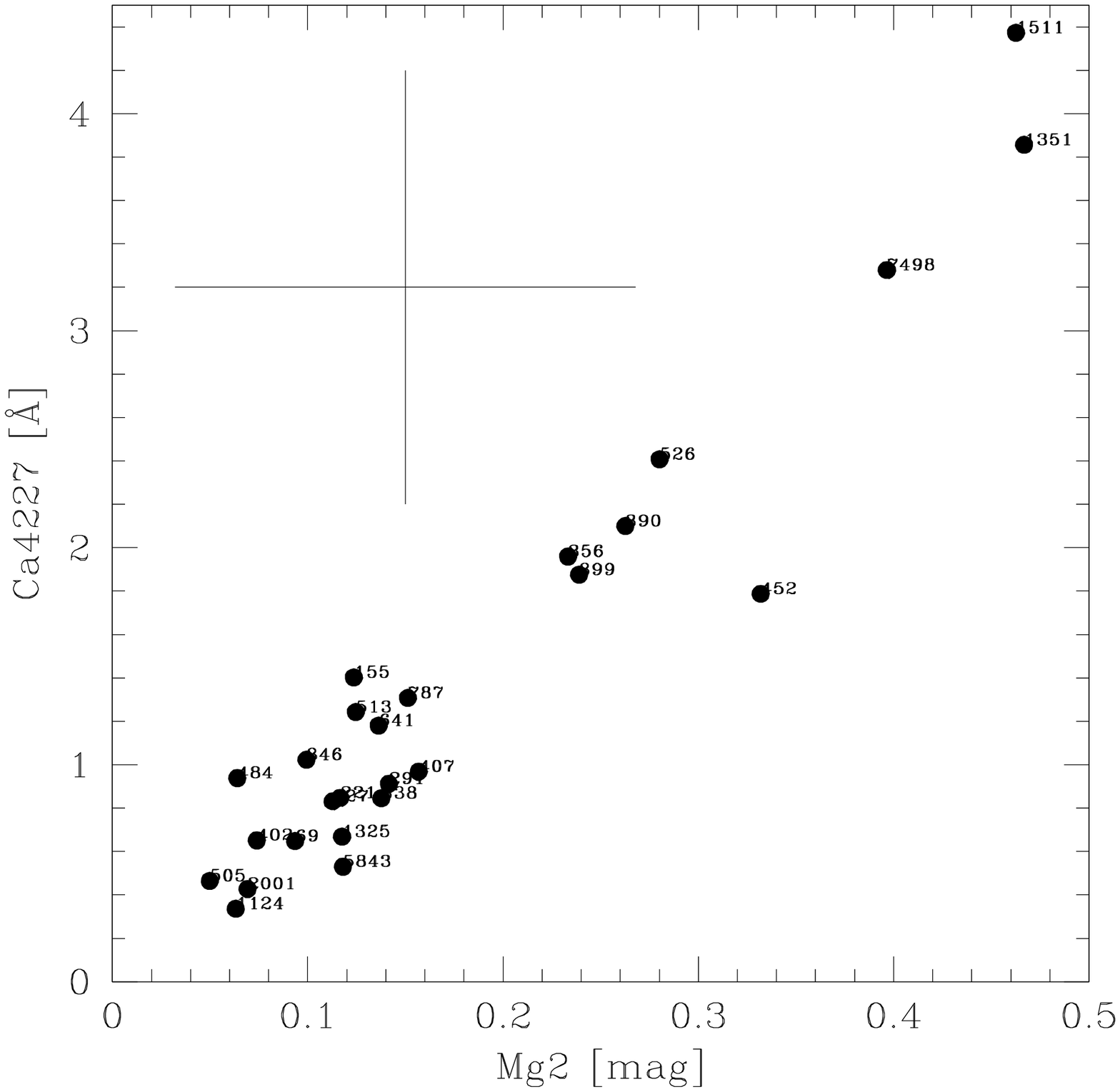}
\caption{{\it Upper panel:} Ca4227 vs. $\langle$Fe$\rangle$, and {\it Lower panel:}
Ca4227 vs. Mg$_2$. {\rm $\langle$Fe$\rangle$ = (Fe4383 + Fe5270 + Fe5335 +
Fe5406)/4}.}
\label{indcafemg2}
\end{figure}

 Figures \ref{indcatemp} and  \ref{indcafemg2} show the behaviour of
 Ca4227 as a function of T$_{\rm eff}$ and the metallicity indicators 
 $\langle$Fe$\rangle$ and Mg$_2$ indices, respectively.
By comparing Fig. \ref{indmg2cntemp} with Fig.  \ref{indcatemp}, a
similarity between the general behaviour of Ca4227 and Mg$_2$ indices with 
T$_{\rm eff}$ is seen.
This explains the behaviour presented between Ca4227 and Mg$_2$ in Fig.
\ref{indcafemg2}.
All these findings show that the cooler giants with higher
metallicity indicators also show high Ca4227 indices. 
These figures also show that the metallicity indices  Mg$_2$ and
$\langle$Fe$\rangle$ grow non-linearly at low temperatures.

\subsection{H$_{\beta}$}

\begin{figure}
\centering
\includegraphics[width=8cm]{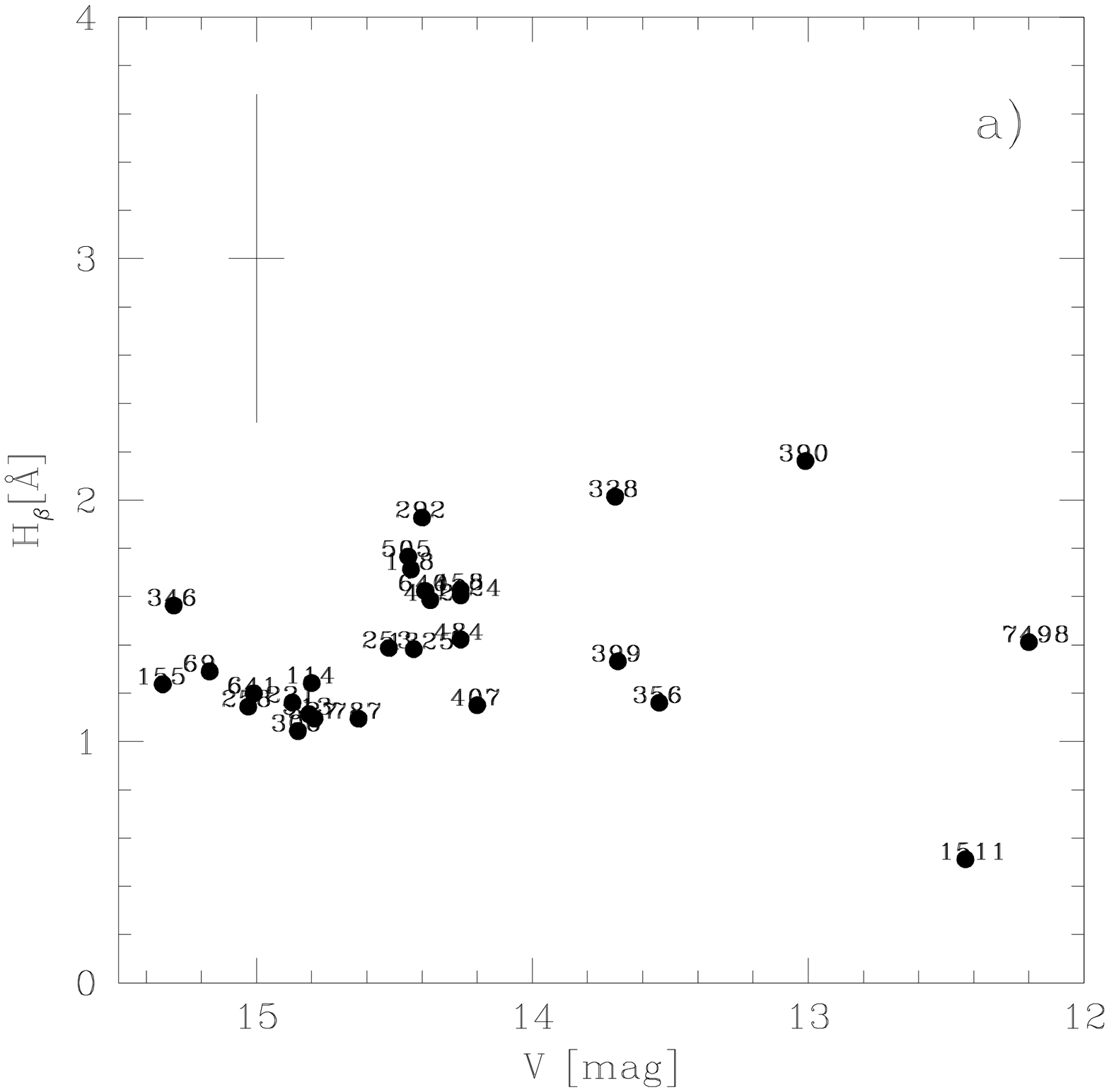}
\includegraphics[width=8cm]{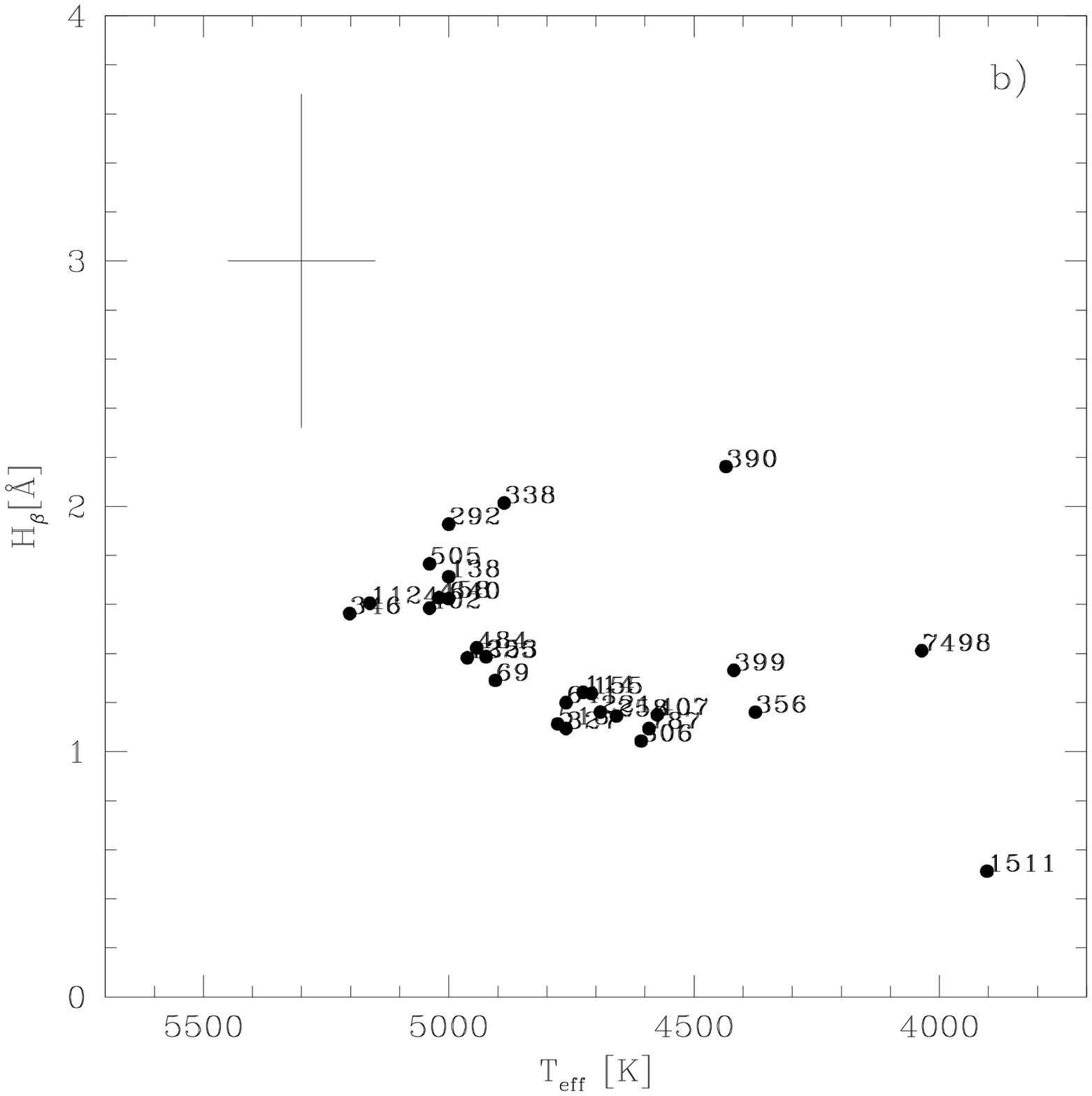}
\caption{{\it a):} H$_{\beta}$ plotted against V, and  {\it b):} H$_{\beta}$
against T$_{\rm eff}$}
\label{indhbetatemp}
\end{figure}

H$_{\beta}$ index is widely used as an age indicator in stellar
population studies and is known to be extremely temperature- and
gravity-dependent. 
We found a mean H$_{\beta}$ index of $\langle$H$_{\beta}$$\rangle$ = 1.57 $\pm$ 0.68
$\rm \AA$ (N = 31). We notice that there is a large variation of H$_{\beta}$ measurements
for stars with V $<$ 14 and this effect appears to be real since 
Figs.\ref{indhbetatemp}a,b suggest  a higher dispersion on the  H$_{\beta}$
measurements for stars cooler than 4500 K.  

\subsection{NaD and Al3953}

NaD and Al3953 features included in the present analysis are as defined in
Worthey et al. (1994) and Serven et al. (2005), respectively. As shown in Table
2 of Serven et al. (2005), the new index Al3953 has a high spectral response to
aluminum abundance variations.
However, although both are very useful indices able to rule out important clues
for understanding the abundance variations in stellar populations, they are
composed of resonance lines and subject to effects from interstellar
absorption, that have to be taken into account when trying to interpret
them.
We found $\langle$NaD$\rangle$ = 4.60 $\pm$ 2.38 $\rm \AA$ (N = 28) and
$\langle$Al3953$\rangle$ = 2.76$\pm$ 1.35 $\rm \AA$ (N = 25). 

\begin{figure}
\centering
\includegraphics[width=8cm]{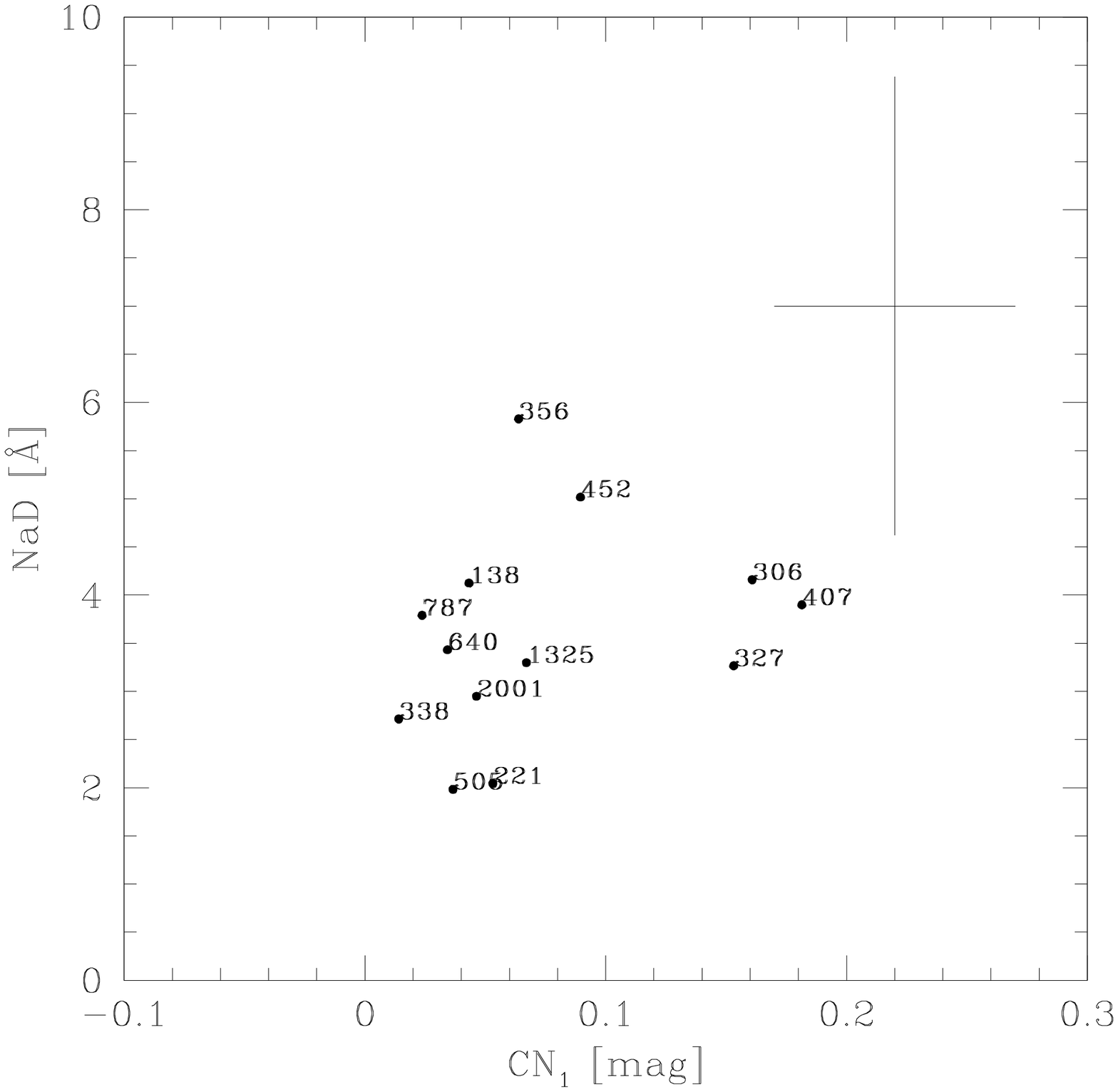}
\includegraphics[width=8cm]{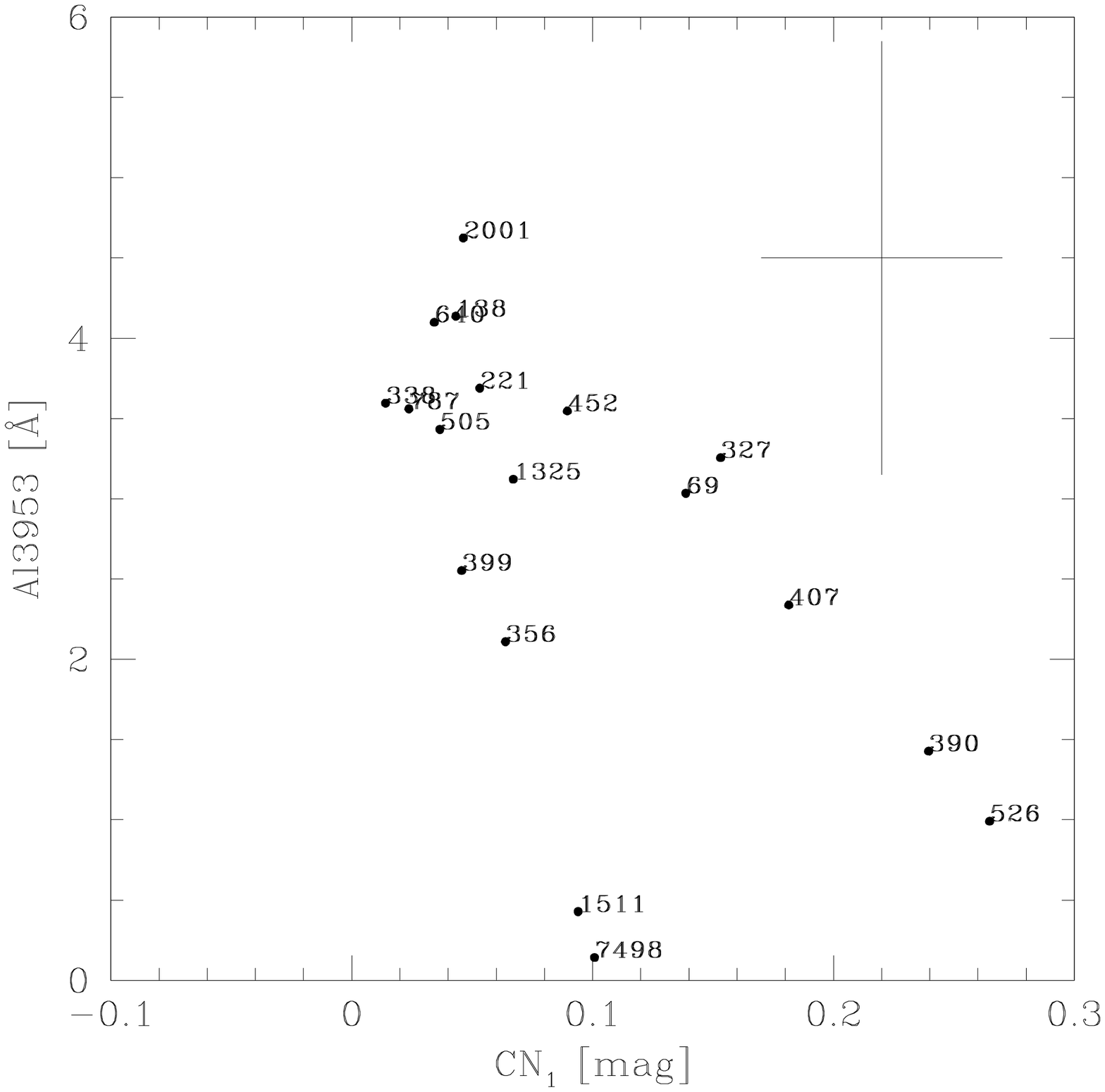}
\caption{NaD and Al3953 as a function of the CN index.}
\label{indnacnal}
\end{figure}

\begin{figure}
\centering
\includegraphics[width=8cm]{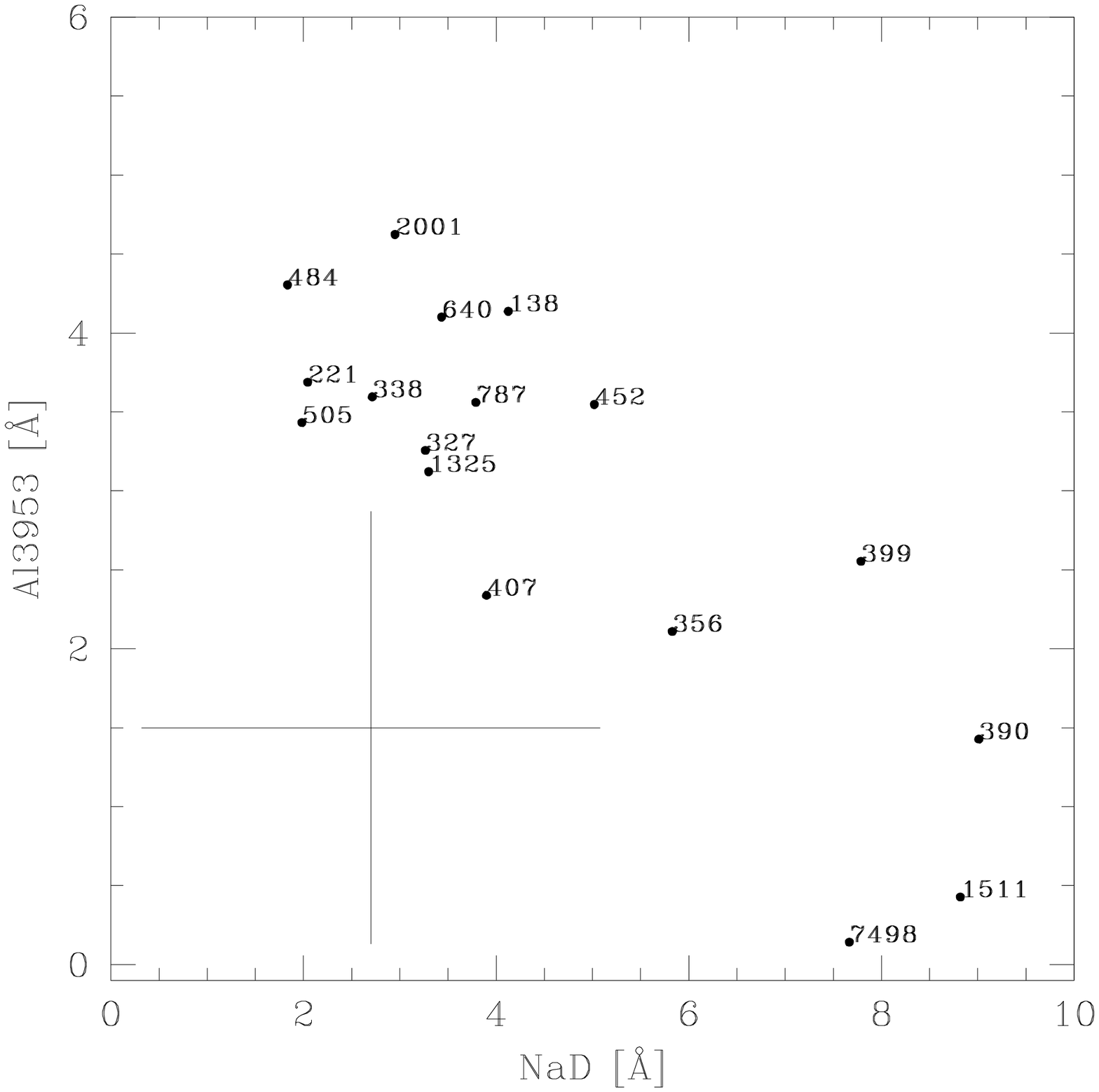}
\includegraphics[width=8cm]{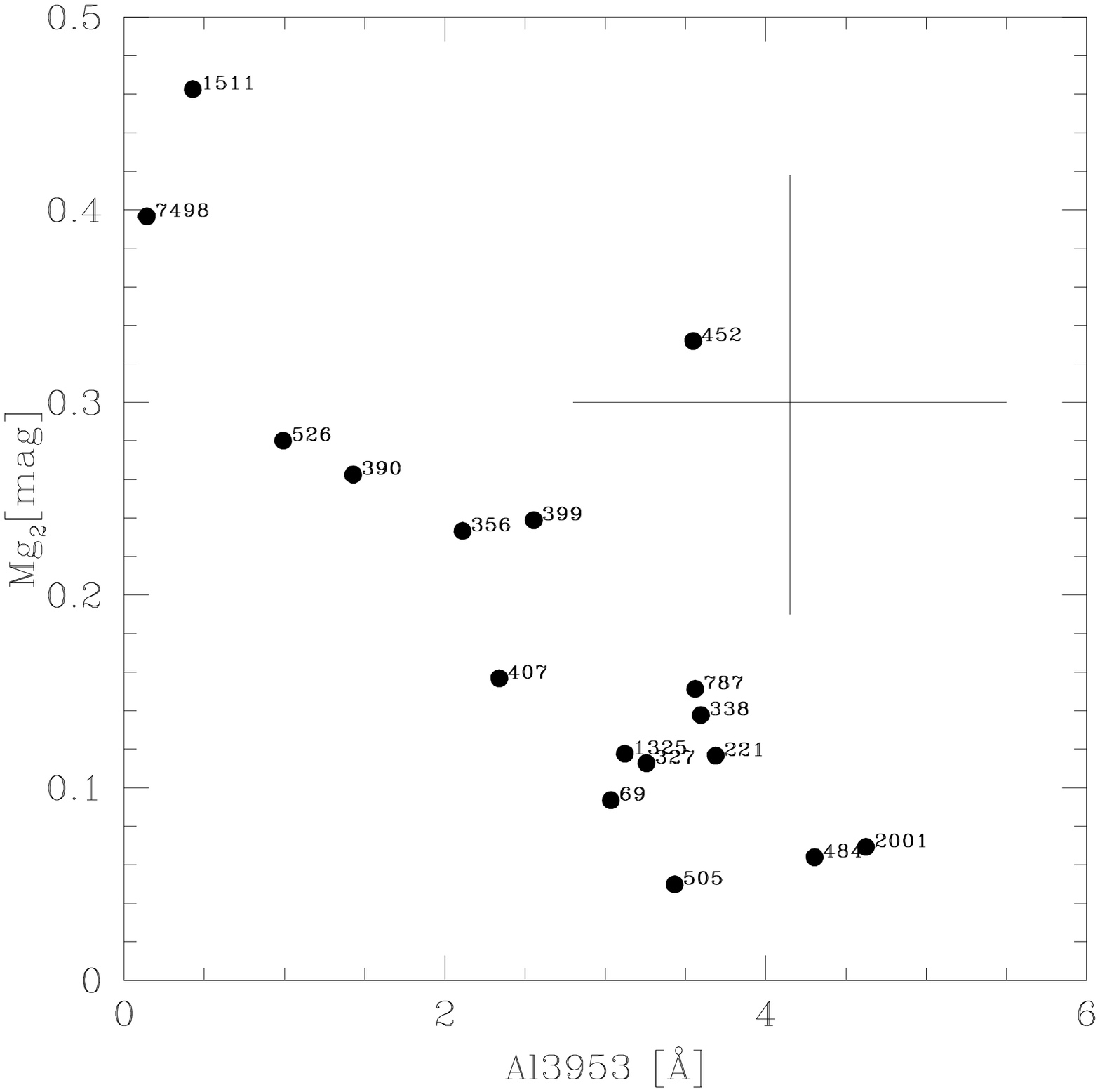}
\caption{{\it Top:} Al3953 vs. NaD. {\it Bottom:} Mg$_{2}$ vs. Al3953.}
\label{indalnamg}
\end{figure}

Figure \ref{indnacnal} displays the behaviour of NaD and
Al3953 as a function of the CN band, while Fig.
\ref{indalnamg} shows the behaviour of NaD and Mg$_{2}$ indices  against the new
index Al3953. An Na-Al anticorrelation is clearly seen.

\subsection{Fitting functions}

The fitting functions approach is a useful tool at low resolution to derive
abundances that are both reddening and distance
modulus independent.  In this section, we use the fitting
functions derived by Barbuy et al. (2003) for the indices Fe5270,
Fe5335 and Mg$_2$, as a function of stellar parameters, in order
to estimate metallicities for M~71 stars, based on the Lick index
measurements presented in the previous sections.

The Fe5270, Fe5335 and Mg$_2$ features are known to be suitable
metallicity indicators due to their relatively weak dependence on
gravity and temperature.  A relation between these indices and
[Fe/H] was derived by Barbuy et al. (2003) based on synthetic spectra
for  resolutions of 8.3 $\rm \AA$ and 3.5 $\rm \AA$.  We use their
Value1 fitting function coefficients for Fe5270, Fe5335 and Mg$_2$
computed with FWHM = 3.5$\rm \AA$, valid for $-$3 $\leq$ [Fe/H]
$\leq$ +0.3, 0 $\leq$ log g $\leq$ 3 and 4000 $\leq$ T$_{\rm eff}$
$\leq$ 7000.  

A mean Fe = (Fe5270+Fe5335)/2 and Mg$_2$ indices were used  as
metallicity indicators.  Mean values of $\langle$Fe5270$\rangle$ =
2.70$\pm$1.47 (N=30), $<$Fe5335$>$ = 2.53$\pm$1.48 (N=29) and
$\langle$Mg$_2$$\rangle$  = 0.17227 $\pm$ 0.11828 (N = 28) were
obtained for giants.  Based on stars in the range 4500 $<$ T$_{\rm
eff}$ $<$ 5000 K (validity of the fitting functions for Mg$_2$, 
mean values of Fe5270 = 2.21$\pm$0.72 (18 giants),
Fe5335 = 2.04$\pm$0.63 (17 giants) and Mg$_2$ = 0.12$\pm$0.03 (11
giants) are found. 

Adopting mean parameters T$_{\rm eff}$ = 4768 K and log g = 2.29
dex, metallicities of [Fe/H]=$-$0.81, $-$0.91 and $-$0.68, corresponding
to Fe5270, Fe5335 and Mg$_2$ respectively, are obtained.
In Table \ref{met}  are reported [Fe/H] values from high-resolution 
spectroscopy and those from fitting functions, where a good
agreement is found.

\begin{table*}
\caption{Literature and present values of metallicity for M71}             
\label{met}      
\centering                          
\begin{tabular}{lcccc}      
\hline\hline                
[Fe/H] & [FeI/H] & [FeII/H] & Method  & Reference                                          \\   
\noalign{\vskip 0.1cm}
\hline                       
$-$0.79 & $-$0.79 & $-$0.79  & High-resolution : 10 giants  	    & Sneden et al. (1994)	    \\
$-$0.73 & ---   &  ---   & Compilation			    & Harris (1996)		    \\
$-$0.71 & $-$0.71 & $-$0.84  & High-resolution : 25 stars		    & Ramirez et al. (2001)	    \\
$-$0.82 & $-$0.81 & $-$0.82  & Revision of Sneden et al.'s 10 giants & Kraft \& Ivans (2003)	    \\
$-$0.80 & $-$0.81 & $-$0.85  & High-resolution : 5 dwarfs		    & Boesgaard et al. (2005)	    \\
$-$0.81 &  ---  & ---    & Fitting Functions		    & Fe5270 index (Present work)   \\
$-$0.91 &  ---  & ---    & Idem				    & Fe5335 index (Present work)   \\
$-$0.68 &  ---  & ---    & Idem				    & Mg$_2$ index (Present work)   \\
\hline                                 
\end{tabular}
\end{table*}

\subsection{Spectrum synthesis}

By adopting [Fe/H] $\approx$ $-$0.8 for M71, taking into account
the values given in Table \ref{met},
we carried out a spectrum synthesis calculation of CN and CH bands
in the wavelength region 4000-4400 {\rm \AA}.
The code is described
by Cayrel et al. (1991) and Barbuy et al. (2003), where CN and CH line lists
are from Kurucz (1993), and implemented in our code as described in
Castilho et al. (1999). MARCS model atmospheres from Gustafsson et al. 
(2006) were used.

\begin{figure}
\centering
\includegraphics[width=8cm]{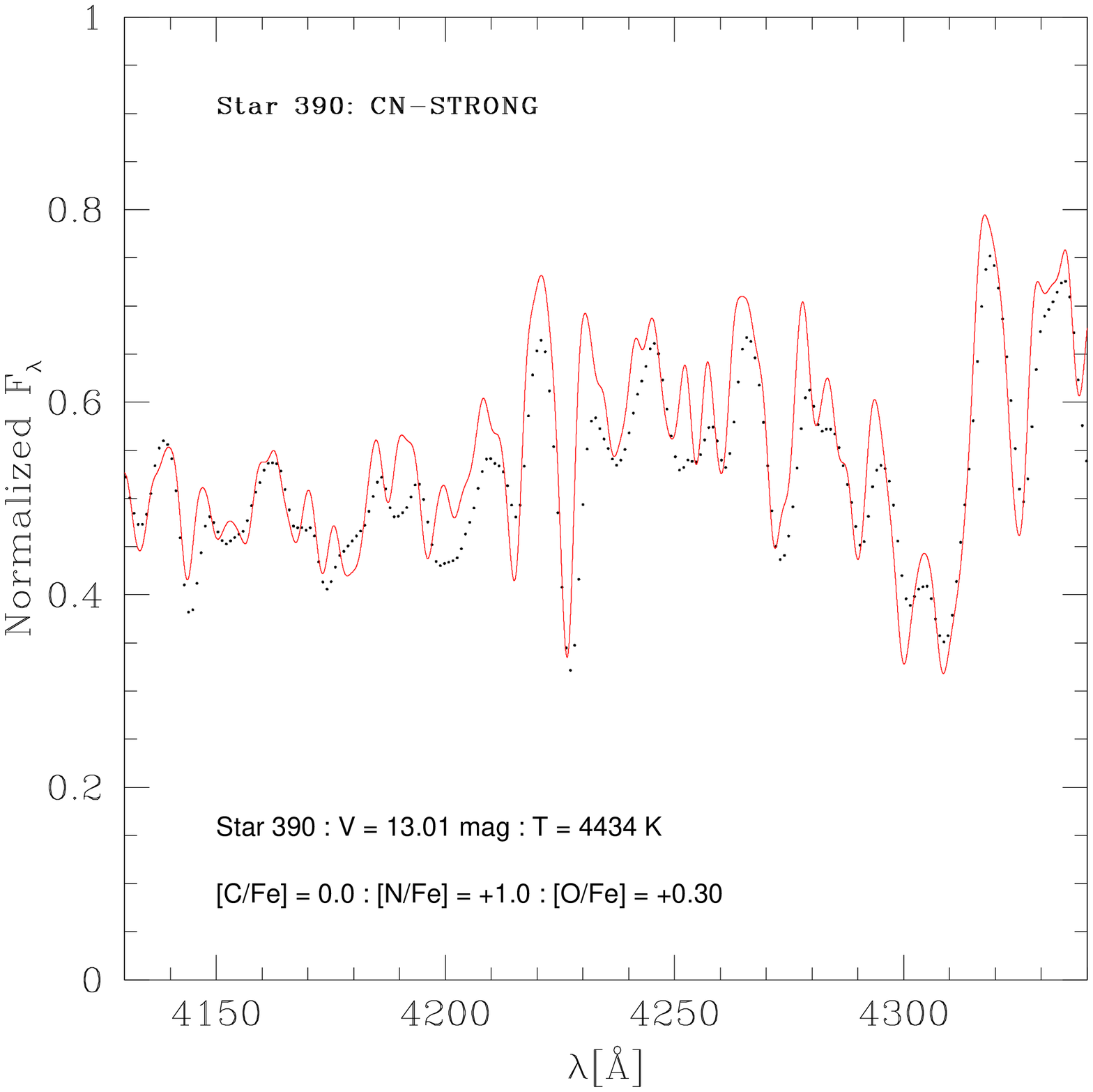}
\includegraphics[width=8cm]{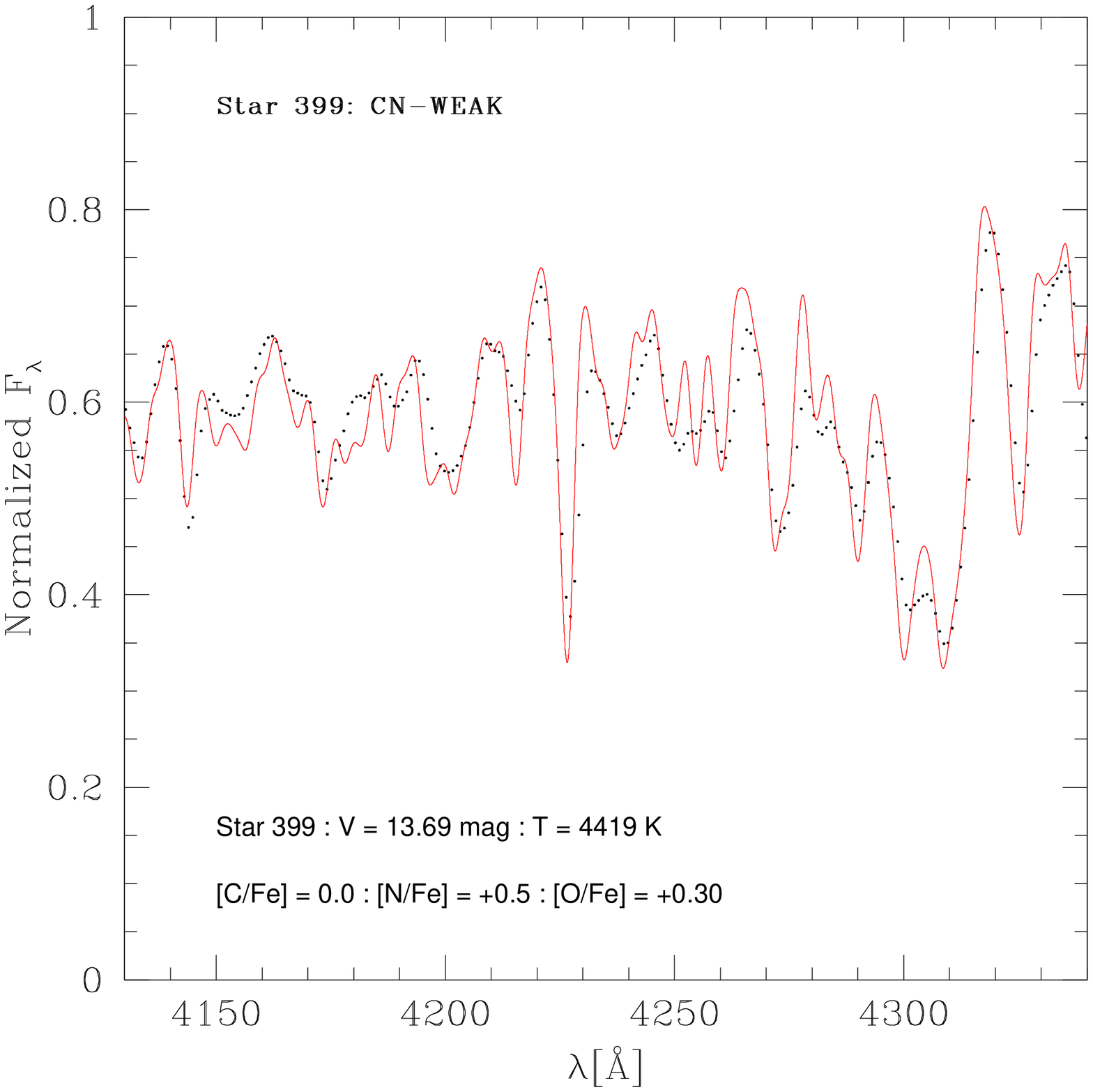}
\caption{Synthetic (solid line) and observed spectra (dotted
line) in the region of CN and CH bands.  {\it Upper panel:} Star
390 (CN-strong): T$_{\rm eff}$, log g, v$_{\rm t}$) = (4434 K, 1.54,
1.5 km/s) computed for [C/Fe]=0.0, [N/Fe]=+1.0; {\it Lower panel:}
Star 399 (CN-weak): (4419 K, 1.8, 1.5 km/s) computed for [C/Fe]=0.0,
[N/Fe]=+0.50 }
\label{cnforte}
\end{figure}

Synthetic spectra were computed for a pair of CN-weak/CN-strong stars with
similar stellar parameters: stars 390 and 399, with 
(T$_{\rm eff}$, log g, v$_{\rm t}$) = (4434 K, 1.54, 1.5 km/s) and (4419 K, 1.8, 1.5 km/s). 
  Carbon and
nitrogen abundances were found to be of [C/Fe]=0.0, [N/Fe]=+1.0 for
the CN-strong star 390
and [C/Fe]=0.0, [N/Fe]=+0.50 for the CN-weak star 399, assuming [O/Fe]=+0.3 in
both cases.
The difference in nitrogen abundances is clear, as shown in Fig.
\ref{cnforte}.

\section{Discussion}

The light elements C, N, O, Na, Mg and Al show variations in globular
cluster stars.  In M71 a bimodal distribution of CN-strong and
CN-weak stars and an anticorrelation between CN and CH were first
made evident by Smith \& Norris (1982).  For stars brighter than
the Horizontal Branch (HB) level they showed that CN and CO are
anticorrelated,  and CN correlated with Na.  Smith \& Penny (1989)
measured CN and CH  based on the bandheads CN 3883 {\rm \AA}, CN
4215 {\rm \AA} and CH 4300 {\rm \AA} for a sample of HB stars in
M71, and revealed an anticorrelation between CN and CH.  CN-strong
and CN-weak components were also revealed by Penny et al. (1992),
who also showed an anticorrelation CN-CH for subgiants in M71.  
A CN-bimodality and a CN-CH anticorrelation  were also found for a
sample of 79 MS stars in M71 by Cohen (1999). Briley et al. (2001)
further derived CN and CH for 75 giants of M71 based on DDO photometric
indices C(41-42) and C(42-45), and confirmed the C vs. N anticorrelation
and the CN-strong and CN-weak components.  Lee (2005) studied CN
and CH band strength variations in 14 M71 giants showing clearly
the CN bimodality, whereas the CN-CH anticorrelation is not clear
in his analysis.

The behaviour of the CN and CH bands above is also reported for
stars on the RGB and MS of 47 Tuc (see e.g. Norris \& Freeman
1979; Norris, Freeman \& Da Costa 1984; Cannon et al. 1998), another
well studied Galactic globular cluster with metallicity close to
that of M71 ([Fe/H] = $-$0.67, Alves-Brito et al. 2005). However,
47 Tuc is  more massive than M~71, as well as  more
concentraded with a half-mass radius r$_{\rm h}$ = 2.79 arcmin and concentration
parameter c=2.03, whereas M~71 has r$_{\rm h}=$ = 1.65 arcmin and c=1.15
(Harris 1996).

 Based on the Cudworth (1985; 2006) star designations, it was
possible to identify several stars in common with other spectroscopic
studies (e.g. in Smith \& Penny 1989; Penny et al. 1992; Lee
2005), which allows us to do a direct CN-CH band-strength
comparison.  The cross-check is possible for 15 out of 22 stars
overlapping with our sample --- 4 stars have no CN measurements in
the literature, whereas  we have no CN measurements for 3 other
stars due to the position of the CCD gaps in their spectra. 
 The comparison shows that 13
out of  15 stars present the same classification of CN-weak or CN-strong
found in the present work, while 2 stars were found to present a
different classification --- stars 1-43 and 1-88 were previously
classified as CN-strong.  On the other hand, within a 
1$\sigma$-uncertainty, those authors
present the same CN values.

Regarding our CN-CH strength results,  Fig. 
\ref{indch} illustrates the CN bimodality and CN-CH anticorrelations.
Note, however, as discussed above, that 3 stars (IDs 1556, 1351 and
640) have inaccurate CN measurements due to the CCD gaps. 
Stars 2001, 458, and 505,  which are classified as CN-weak,
show lower CH values than other CN-weak stars.
Even with some statistical outliers our CN-CH distribution has similar
proportions as those given in Smith \& Norris (1982).  

Within the canonical framework of stellar structure and evolution of
 low-mass stars, 
the enhancement of nitrogen followed by the depletion of carbon
in the CN-strong stars is attributed to the dredge-up of CN-processed material 
into the surface layers of such stars (Iben 1964; Charbonnel 1994). 
The present CN variations followed by an anticorrelation in the CH band 
confirm  that an episode of deep mixing occurred in the M71 giants studied.
On the other hand, the CN bimodality is better explained within a primordial
pollution scenario.
The mixing hypothesis does not explain the behaviour 
seen for the positive correlation between CN-Na and CN-Al,
given that Na and Al are not produced during these mixing events.

Elements such as Al and Mg are produced by p-capture at higher
temperatures (e.g. Langer \& Hoffman 1995). As shown in Fig.
\ref{indalnamg}, there is a clear anticorrelation between Al3953
vs. NaD and Mg$_2$ indices.  The Al-Mg anticorrelation seen in some
globular cluster red giants is well explained by the enhanced extra
mixing discussed in Denissenkov \& VandenBerg (2003). Ramirez \& Cohen (2002) reported a [Na/Fe]:[Al/Fe]
correlation for their sample of M71 stars, however their results
show a  2$\sigma$ uncertainty (see their Fig. 13).

 A scenario that appears plausible is that CNO-processed material
in intermediate mass stars (IMS) on the AGB  occurs
early in the cluster's life, and their stellar winds are captured by low-mass
stars (Gratton et al. 2004) or else by fast rotating massive stars
(Decressin et al. 2007). Both the AGB or massive star early pollution
possibilities are consistent with scenarios of self-enrichment
proposed by Cayrel (1986), Parmentier et al. (1999),
Parmentier \& Gilmore (2001), Thoul et al. (2002), and Bekki et al. (2007).
The latter developed a model of globular cluster formation
in the central regions of low mass proto-galaxies embedded in dark matter
halos. These proto-galaxies would retain the AGB ejecta and cause
 an ``external pollution'' of the globular cluster stars.
This model can explain the C-N and Mg-Al anticorrelations, but
shows a strong disagreement with the observed O-Na anticorrelation.

For the present sample, we found that CN and Na are correlated, while a
CN-Al correlation does not appear. These CN-Na results imply a primordial
explanation for abundance variations in M71.

\section{Summary and conclusions}

In M71 previous studies such that of Smith \& Norris (1982) that first showed
 a bimodal distribution of CN-strong and CN-weak
stars and an anticorrelation between CN and CH. This was followed by
Smith \& Penny (1989), Penny et al. (1992), Briley et al. (2001),
Cohen (1999), Ramirez et al. 2001, Ram\'{\i}rez \&
Cohen (2002), Boesgaard et al. (2005), and  Lee (2005).

We measured CN, CH, Ca4277, iron
and magnesium indicators, H$_{\beta}$, NaD and Al3953 spectral indice, from
 low-resolution spectra of 89 stars
of the metal-rich globular cluster M71, observed with the Gemini Multi-Object 
Spectrograph (GMOS) at the Gemini-North telescope.
 CN and CH strengths were obtained for 89 stars,
among which 33 giants.
As expected from evolutionary mixing theories and additional
extra-mixing (Iben 1964; Charbonnel 1994; Denissenkov \& VandenBerg 2003),
we find a CN-CH anticorrelation.
We find CN-strong and CN-weak stars, with around 30\% of CN-strong ones,
similar to other clusters such as NGC 6752 with about 50\%.

We confirm a CN-bimodality besides the CN-CH anticorrelation, a CN-Na correlation,
and  Al-Na and Mg$_2$-Al anticorrelation.
The interpretation of CN bimodality is instead better understood
in terms of primordial variations, and possible scenarios include
 an early enrichment by winds from intermediate mass stars in the AGB phase,
and captured by low-mass stars, early in the cluster's life (Gratton et al.
2004 and references therein; Bekki et al. 2007) or early pollution by fast
rotating massive stars (Decressin et al. 2007).

 CN-strong and CN-weak bimodality is only seen in relatively metal-rich
globular clusters, but not in all of them. Such behaviour is well studied
in particular in 47 Tucanae, NGC 6752 and M4.
Such abundance variations in metal-rich globular clusters  
is undoubtedly one of the most intrincate challenges for the current theory of
stellar evolution, and further observations with larger samples would
be interesting.

\begin{acknowledgements}

AA acknowledges a FAPESP fellowship no. 04/00287-9.  AA would
like to thank the hospitality of the Department of Astronomy at the
University of Virginia, during a visit in which part of this work
was developed.  Likewise, RPS would like to warmly than the hospitality
of the Astronomy Department at the University of S\~ao Paulo, during
which much of this work was conceived.  We are grateful to Professor
Kyle Cudworth for cordially  making his preliminary reduction
of proper motions and photometry for M71 stars available to us.
All observations were part of GEMINI Program GN-2002B-Q-42.  This
work is based on observations obtained at the Gemini Observatory,
which is operated by the AURA, Inc., under a cooperative agreement
with the NSF on behalf of the Gemini partnership: the NSF (United
States), the PPARC (United Kingdom), the NRC (Canada), CONICYT
(Chile), the ARC (Australia), CNPq (Brazil) and CONICET (Argentina)

\end{acknowledgements}

\Online

\end{document}